\documentclass[fleqn,usenatbib]{mnras}
\usepackage{newtxtext,newtxmath}
\usepackage{fontenc}
\usepackage{ae,aecompl}
\usepackage{footnote}
\usepackage{cleveref}

\usepackage{graphicx}
\usepackage{amsmath}

\newcommand{\msun}{\hbox{M$_{\odot}$}}

\newcommand{\halpha}{\hbox{H$\alpha$ }}
\newcommand{\hbeta}{\hbox{H$\beta$ }}

\newcommand{\swift}{\textit{Swift }}

\title[Discovery of the TDE ASASSN-19dj]{Discovery and follow-up of ASASSN-19dj: An X-ray and UV luminous TDE in an extreme post-starburst galaxy}

\author[Hinkle et al.]
{\href{http://orcid.org/0000-0001-9668-2920}{Jason T. Hinkle}$^{1}$\thanks{E-mail: jhinkle6@hawaii.edu},
\href{http://orcid.org/0000-0001-9206-3460}{T. W.-S. Holoien}$^{2}$\thanks{Carnegie Fellow}, 
\href{https://orcid.org/0000-0002-4449-9152}{K. Auchettl}$^{14,15,3,4}$, 
\href{https://orcid.org/0000-0003-4631-1149}{B. J. Shappee}$^{1}$, 
\newauthor
\href{https://orcid.org/0000-0001-7351-2531}{J.~M.~M.~Neustadt}$^{5}$, 
\href{https://orcid.org/0000-0003-3490-3243}{A.~V.~Payne}$^{1}$\thanks{NASA Fellowship Activity Fellow},
J.~S.~Brown$^{4}$, 
\href{https://orcid.org/0000-0001-6017-2961}{C.~S.~Kochanek}$^{5,6}$,
K.~Z.~Stanek$^{5,6}$, 
\newauthor
M.~J.~Graham$^{7}$, 
\href{https://orcid.org/0000-0002-2471-8442}{M. A. Tucker}$^{1}$\thanks{DOE CSGF Fellow},
A. Do$^{1}$,
\href{https://orcid.org/0000-0003-0227-3451}{J.~P. Anderson}$^{8}$,
S. Bose$^{5,6}$,
P. Chen$^{9}$,
\newauthor 
D.~A.~Coulter$^{4}$, 
G. Dimitriadis$^{4}$, 
\href{https://orcid.org/0000-0002-1027-0990}{Subo Dong}$^{9}$, 
R.~J.~Foley$^{4}$,
M. E. Huber$^{1}$,
T. Hung$^{4}$,
\newauthor 
C.~D.~Kilpatrick$^{4}$, 
G. Pignata$^{11,10}$,
A. L. Piro$^{2}$,
\href{https://orcid.org/0000-0002-7559-315X}{C.~Rojas-Bravo}$^{4}$, 
M. R. Siebert$^{4}$,
\newauthor
B. Stalder$^{12}$,  
\href{http://orcid.org/0000-0003-2377-9574}{Todd A.~Thompson}$^{5,6}$,
J. L. Tonry$^{1}$,
P. J. Vallely$^{5}$, \&
\href{http://orcid.org/0000-0001-9209-1808}{J.~P.~Wisniewski}$^{13}$ \\
$^{1}$Institute for Astronomy, University of Hawai`i, 2680 Woodlawn Dr., Honolulu, HI 96822, USA\\
$^{2}$The Observatories of the Carnegie Institution for Science, 813 Santa Barbara St., Pasadena, CA 91101, USA\\
$^{3}$DARK, Niels Bohr Institute, University of Copenhagen, Lyngbyvej 2, 2100 Copenhagen, Denmark\\
$^{4}$Department of Astronomy and Astrophysics, University of California, Santa Cruz, CA 95064, USA\\
$^{5}$Department of Astronomy, The Ohio State University, 140 West 18th Avenue, Columbus, OH 43210, USA\\
$^{6}$Center for Cosmology and Astroparticle Physics, The Ohio State University, 191 W.~Woodruff Avenue, Columbus, OH 43210, USA\\
$^{7}$California Institute of Technology, 1200 E. California Blvd, Pasadena, CA 91125, USA\\
$^{8}$European Southern Observatory, Alonso de C{\'o}rdova 3107, Vitacura, Casilla 19001, Santiago, Chile\\
$^{9}$Kavli Institute for Astronomy and Astrophysics, Peking University, Yi He Yuan Road 5, Hai Dian District, Beijing 100871, China\\\
$^{10}$Millennium Institute of Astrophysics, Santiago, Chile\\
$^{11}$Departamento de Ciencias Fisicas, Universidad Andres Bello, Avda. Republica 252, Santiago, Chile\\
$^{12}$Rubin Observatory Project Office, 950 North Cherry Avenue, Tucson, AZ 85719, USA\\
$^{13}$Homer L. Dodge Department of Physics \& Astronomy, The University of Oklahoma, 440 W. Brooks Street, Norman, OK 73019, USA\\
$^{14}$ School of Physics, The University of Melbourne, Parkville, VIC 3010, Australia\\
$^{15}$ ARC Centre of Excellence for All Sky Astrophysics in 3 Dimensions (ASTRO 3D), Australia
}

\date{Accepted 2020 October 8. Received 2020 September 30; in original form 2020 June 18}
\pubyear{2020}

\begin{document}
\label{firstpage}
\pagerange{\pageref{firstpage}--\pageref{lastpage}}
\maketitle

\begin{abstract}
We present observations of ASASSN-19dj, a nearby tidal disruption event (TDE) discovered in the post-starburst galaxy KUG 0810+227 by the All-Sky Automated Survey for Supernovae (ASAS-SN) at a distance of d $\simeq98$ Mpc. We observed ASASSN-19dj from $-$21 to 392 d relative to peak ultraviolet (UV)/optical emission using high-cadence, multiwavelength spectroscopy and photometry. From the ASAS-SN $g$-band data, we determine that the TDE began to brighten on 2019 February 6.8 and for the first 16 d the rise was consistent with a flux $\propto t^2$ power-law. ASASSN-19dj peaked in the UV/optical on 2019 March 6.5 (MJD = 58548.5) at a bolometric luminosity of $L = (6.2 \pm 0.2) \times 10^{44} \text{ erg s}^{-1}$. Initially remaining roughly constant in X-rays and slowly fading in the UV/optical, the X-ray flux increased by over an order of magnitude $\sim$225 d after peak, resulting from the expansion of the X-ray emitting region. The late-time X-ray emission is well fitted by a blackbody with an effective radius of $\sim1 \times 10^{12} \text{ cm}$ and a temperature of $\sim6 \times 10^{5} \text{ K}$. The X-ray hardness ratio becomes softer after brightening and then returns to a harder state as the X-rays fade. Analysis of Catalina Real-Time Transient Survey images reveals a nuclear outburst roughly 14.5 yr earlier with a smooth decline and a luminosity of $L_V\geq1.4 \times 10^{43}$ erg s$^{-1}$, although the nature of the flare is unknown. ASASSN-19dj occurred in the most extreme post-starburst galaxy yet to host a TDE, with Lick H$\delta_{A}$ = $7.67\pm0.17$ \AA.
\end{abstract}

\begin{keywords}
accretion, accretion discs --- black hole physics --- galaxies: nuclei
\end{keywords}

\section{Introduction}
Supermassive black holes (SMBHs) are known to reside in the centres of most massive galaxies \citep[e.g.][]{kormendy95, magorrian98, rees88, ho08, gultekin09, kormendy13}. If mass is actively accreting on to these SMBHs, they can be detected as Active Galactic Nuclei (AGNs). Conversely, direct detections of inactive SMBHs are difficult, mainly limited to our own black hole \citep[Sgr A*;][]{ghez05}, or massive ($\ga 10^6 \ \msun$) SMBHs in nearby ($\la \ 50$ Mpc) galaxies, where stars \citep[e.g.][]{kormendy96, gebhardt11} and/or gas \citep[e.g.][]{ford94, atkinson05} within the SMBH's sphere of influence can be resolved. Only one SMBH, P\={o}wehi in M87, has been directly observed, by the Event Horizon Telescope \citep[][]{eht19}. Tidal disruption events (TDEs) provide an opportunity to study otherwise inactive SMBHs at greater distances.

A TDE occurs when a star passes within the tidal radius of a SMBH and is torn apart, resulting in a luminous accretion flare \citep[][]{rees88, phinney89, evans89, ulmer99, komossa15, stone19}. Early theoretical work predicted that the blackbody temperatures of TDEs should be on the order of $10^5$ K, consistent with a peak in the soft X-ray band \citep[e.g.][]{lacy82, rees88, evans89, phinney89}, but observational studies have discovered a breadth of TDE phenomenology. For example, TDE candidates have been detected in the hard X-ray \citep[e.g.][]{bloom11, burrows11, cenko12b, pasham15}, soft X-ray \citep[e.g.][]{bade96, komossa99a,komossa99b,grupe99, auchettl17}, ultraviolet (UV) \citep[e.g.][]{stern04, gezari06, gezari08, gezari09}, optical \citep[e.g.][]{vanvelzen11, gezari12b, cenko12a, arcavi14, chornock14, holoien14a, vinko15, holoien16a, holoien16b, brown18, holoien19b, holoien19c}, and radio \citep[e.g.][]{zauderer11, cenko12b, velzen16, alexander16, alexander17, brown17a}, with many showing emission in multiple energy bands. 

The diversity seen in these events fuelled a broad range of theoretical investigations \citep[e.g.][]{lodato15, krolik16, svirski17, ryu20a, krolik20}. The unifying model of \citet{dai18} may provide an explanation of the diversity, positing that many of the observed multiwavelength photometric and spectroscopic properties of TDEs are a result of the viewing angle. Despite this, the origin of the UV/optical emission is still debated, with reprocessed emission from an accretion disc \citep[e.g.][]{dai18, mockler19} and shocks from stream-stream collisions \citep[e.g.][]{jiang16, bonnerot17, lu20, ryu20a} as the most commonly proposed emission mechanisms. Due to similar energy budgets and the general lack of observational features to neatly disambiguate models for a particular event, the nature of the UV/optical emission remains an open question.

Observations of TDEs may provide information on the physics of accretion \citep[e.g.][]{lodato11, guillochon15, metzger16, shiokawa15}, shock physics \citep[e.g.][]{lodato09}, jet formation \citep[e.g.][]{farrar14, wang16, biehl18}, and the environment and growth of SMBHs \citep[e.g.][]{auchettl18}. However, the characteristics of the observed emission from TDEs, such as their light curves, spectroscopic evolution (both optical and X-ray), blackbody properties, etc., are a function of many physical parameters. Such properties include the star's impact parameter \citep[e.g.][]{guillochon13, guillochon15, gafton19}, mass \citep[e.g.][]{gallegos-garcia18, mockler19, law-smith19}, composition \citep[e.g.][]{kochanek16a}, evolutionary stage \citep[e.g.][]{macleod12}, age \citep[e.g.][]{gallegos-garcia18}, and spin \citep[e.g.][]{golightly19}. Additionally, stellar demographics \citep[e.g.][]{kochanek16b}, the fraction of accreted stellar material \citep[e.g.][]{metzger16, coughlin19}, and the geometry of accretion \citep[e.g.][]{kochanek94, lodato11, guillochon15, metzger16, dai15, shiokawa15, dai18} may affect the observed emission. 

It has also been shown that TDE emission may be sensitive to black hole spin and mass \citep[e.g.][]{ulmer99, graham01, mockler19, gafton19}, making TDEs useful probes of otherwise quiescent SMBHs. As such, TDE light curves can be used to constrain the masses of SMBHs, which are consistent with those derived from other methods \citep{mockler19}. While there are a large number of potentially relevant physical parameters, the observed UV/optical emission is relatively well fitted by a blackbody \citep[e.g.][]{gezari12b, holoien14b, holoien16a, holoien16b, brown16a, hung17, holoien18a, holoien19c, leloudas19, vanvelzen20, holoien20}. It has also been shown that the peak UV/optical luminosities of TDEs are related to their decline rates \citep{mockler19, hinkle20a}, with more luminous TDEs declining more slowly after peak. As the number of TDEs increases, they will provide a more complete picture of SMBH growth and evolution via accretion and the central environments of galaxies.

The spectroscopic properties of optical TDEs are varied \citep[e.g.][]{arcavi14, hung17, leloudas19, wevers19, vanvelzen20, holoien20}, with differences in observed species, line strengths/widths, and line ratios. Emission lines from hydrogen, helium, and more exotic features due to Bowen fluorescence have been observed \citep[e.g.][]{leloudas19, vanvelzen20}. Most TDEs have simple broad line profiles, but others have double-peaked disc-like line profiles \citep[e.g.][]{holoien19b, hung20} or strong narrow lines \citep[e.g.][]{vanvelzen20, holoien20}. Possible explanations for this variety are details in the physics of photoionization \citep[e.g.][]{guillochon14, gaskell14, roth16, kara18, leloudas19}, differences in the composition of stars due to evolution \citep{kochanek16a}, the viewing geometry with respect to an accretion disc \citep[e.g.][]{holoien19b, short20, hung20}, or even the disruption of helium stars \citep{gezari12b, strubbe15}. Additionally, there have been variations in the times at which strong emission lines appear \citep[e.g.][]{holoien19b, holoien19c, holoien20}. An even larger sample of optically bright TDEs will better constrain the mechanisms that influence the observed emission from such events.

TDEs are rare, with an expected frequency between $10^{-4}$ and $10^{-5} \text{ yr}^{-1}$ per galaxy \citep[e.g.][]{vanvelzen14, holoien16a, vanvelzen18, auchettl18}. Interestingly though, TDEs seem to prefer post-starburst host galaxies. In such galaxies, the TDE rates can be enhanced by up to 200 times as compared to the average rates \citep[e.g.][]{arcavi14, french16, law-smith17, graur18}. Combining these suggests that in the most extreme post-starbursts, TDEs can occur at roughly the same rate as other bright transients like supernovae.

There are few observations of the early-time evolution of TDEs, a time period that may be important to understanding how the disrupted stellar material settles into an accretion flow. With the advent of transient surveys like the All-Sky Automated Survey for Supernovae \citep[ASAS-SN;][]{shappee14, kochanek17}, the Asteroid Terrestrial Impact Last Alert System \citep[ATLAS;][]{tonry18}, the Zwicky Transient Facility \citep[ZTF;][]{bellm19}, the Panoramic Survey Telescope and Rapid Response System \citep[Pan-STARRS;][]{chambers16} and the Young Supernova Experiment \citep[YSE;][]{2019TNSAN.148....1J}, many more TDEs are being discovered. This includes an increasing number of TDEs discovered before their peak brightness \citep[e.g.][]{holoien19b, holoien19c, leloudas19, vanvelzen19, wevers19, vanvelzen20, holoien20}.

While these fast-cadence, wide-field optical surveys are ideal for discovering TDEs, a significant fraction of emission from some events is in the soft X-ray band \citep[e.g.][]{ulmer99, auchettl17}. Recently, an increasing number of TDE candidates discovered in the optical have exhibited strong X-ray emission. Examples include ASASSN-14li \citep[e.g.][]{miller15, holoien16a, brown17a}, ASASSN-15oi \citep[e.g.][]{holoien16c, gezari17, holoien18a}, ASASSN-18ul \citep[][; Payne et al., in preparation]{wevers19}, Gaia19bpt \citep{vanvelzen20}, ZTF19aapreis \citep{vanvelzen20}, and the TDE studied in this work, ASASSN-19dj. The combination of UV, optical and X-ray has given greater insight on the formation of an accretion disc, reprocessing, and the differences between thermal (non-jetted) and non-thermal (jetted) TDEs \citep[e.g.][]{auchettl17}. In general, long-term X-ray light curves of TDE candidates are required to distinguish them from AGNs and to study detailed accretion physics \citep[e.g.][]{auchettl18}.

In this paper we present the discovery and observations of ASASSN-19dj. Smaller data sets on ASASSN-19dj have been analysed by \citet{liu19} and as part of the larger sample in \citet{vanvelzen20}. Here we provide analysis of the host galaxy in addition to a longer observational baseline with corresponding detailed analysis of the UV/optical photometric, optical spectroscopic, and X-ray properties of ASASSN-19dj. Throughout the paper we assume a cosmology of $H_0$ = 69.6 km s$^{-1}$ Mpc$^{-1}$, $\Omega_{M} = 0.29$, and $\Omega_{\Lambda} = 0.71$. This paper is organised as follows. In Section \ref{obs} we detail the discovery and observations of the TDE. In Section \ref{analysis} we present the analysis our of results. Section \ref{sec:disc} provides a discussion of our results. Finally, our analysis is summarized in Section \ref{summary}.

\section{Discovery and Observations}\label{obs}

\begin{figure*}
\centering
 \includegraphics[width=.329\textwidth]{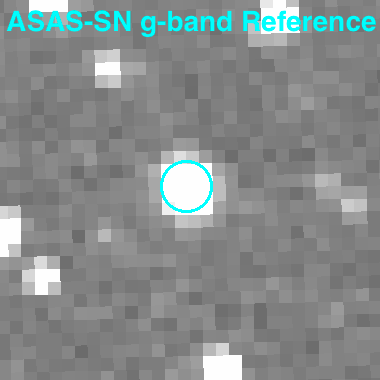}\hfill
 \includegraphics[width=.33\textwidth]{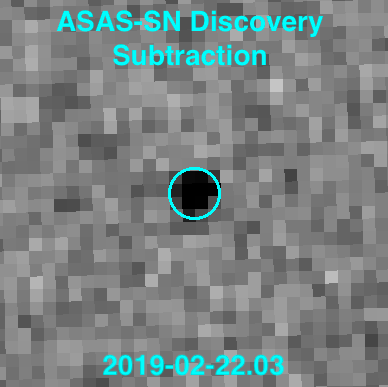}\hfill
 \includegraphics[width=.329\textwidth]{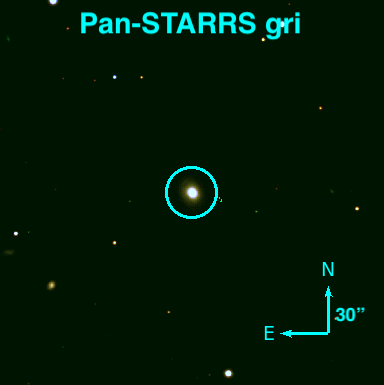}
 \caption{The ASAS-SN $g$-band reference image for the location of ASASSN-19dj (left), the subtracted ASAS-SN $g$-band discovery image from 2019-02-22.03 showing flux from ASASSN19-dj (center), and a combined Pan-STARRS $gri$ colour image of the host galaxy (right). The cyan circle (of radius 15 arcsec) marks the location of ASASSN-19dj.}
 \label{fig:disc_image}
\end{figure*}

ASASSN-19dj $(\alpha,\delta) =$ (08:13:16.96, $+$22:38:54.00) was discovered in the $g$-band in data from the ASAS-SN ``Bohdan Paczy\'nski'' unit in Cerro Tololo, Chile on 2019 February 22 \citep{brimacombe19ATel}. Its discovery was announced on the Transient Name Server (TNS), and assigned the name AT 2019azh\footnote{\url{https://wis-tns.weizmann.ac.il/object/2019azh}}. Rather than anonymise the discovering survey, in this paper, we will continue to refer to the TDE by its survey name ASASSN-19dj. ASASSN-19dj is located in the nucleus of the post-starburst galaxy KUG 0810+227, at a redshift of z = 0.022346 \citep{adelmanmccarthy06}. This redshift corresponds to a luminosity distance of 97.9 Mpc, making ASASSN-19dj one of the closest TDEs discovered to date. The $g$-band reference used for host subtraction, the discovery image of ASASSN-19dj, and a false-colour Pan-STARRS $gri$ image of the host galaxy\footnote{\url{http://ps1images.stsci.edu/cgi-bin/ps1cutouts?pos=123.320605325\%2B22.648343&filter=color}} \citep{chambers16} are shown in Figure \ref{fig:disc_image}. The circle marking the location of ASASSN-19dj is 15 arcsec in radius, the same as the apertures used for the photometry presented in this paper. 

Multiple spectroscopic observations were obtained shortly after discovery. Both the Nordic Optical Telescope Unbiased Transient Survey \citep[NUTS;][]{heikkila19} and the extended Public ESO Spectroscopic Survey for Transient Objects \citep[ePESSTO;][]{barbarino19} obtained spectra that showed a strong blue continuum with few strong spectral features compared to the Sloan Digital Sky Survey \citep[SDSS;][]{york00} host spectrum (see \S \ref{sec:archival}). The strong blue continuum, the appearance of broad H$\alpha$ emission lines, and a position consistent with the nucleus of the host galaxy made ASASSN-19dj a strong TDE candidate. Based on this, we triggered spectroscopic and ground-based photometric (Swope and LCOGT) follow-up of ASASSN-19dj.

Using ZTF and \textit{Neil Gehrels Swift Gamma-ray Burst Mission} \citep[\textit{Swift};][]{gehrels04} observations, \citet{vanvelzen19} observed a plateau in the optical and UV light curve between 2019 February 24.25 and 2019 March 11.45. From fits to the ZTF and \swift photometry, they found that the transient spectral energy distribution (SED) was consistent with a T = $(3.2\pm 0.7) \times 10^4 $~K blackbody, and measured a spatial separation from the host nucleus of 0.07 $\pm$ 0.31 arcsec. They classified the source as a TDE based on the observations of multiple blue spectra, a hot blackbody temperature, position in the centre of the host galaxy, and the lack of spectral features usually associated with AGN or supernovae. Using the central galaxy velocity dispersion from SDSS DR14 and the scaling relationship of \citet{gultekin09}, they calculated a SMBH mass of $M_{BH} \lesssim 4 \times 10^6$~\msun, and suggested that the observed plateau in the light curve was the result of Eddington-limited accretion.

\subsection{Archival Data of KUG 0810+227} \label{sec:archival}
KUG 0810+227 has been observed by several sky surveys across the electromagnetic spectrum. We obtained $ugriz$ and $JHK_S$ images from SDSS Data Release 15 \citep{aguado19} and the Two Micron All-Sky Survey \citep[2MASS;][]{skrutskie06}, respectively. We measured aperture magnitudes using a 15\farcs{0} aperture radius in order to capture all of the galaxy light, and used several stars in the field to calibrate the magnitudes. We also obtained an archival $NUV$ magnitude from the Galaxy Evolution Explorer \citep[GALEX;][]{martin05} All-sky Imaging Survey (AIS) catalog and $W1$ and $W2$ magnitudes from the Wide-field Infrared Survey Explorer \citep[WISE;][]{wright10} AllWISE catalog, giving us coverage from ultraviolet through mid-infrared wavelengths.

\begin{figure*}
\centering
 \includegraphics[width=.48\textwidth]{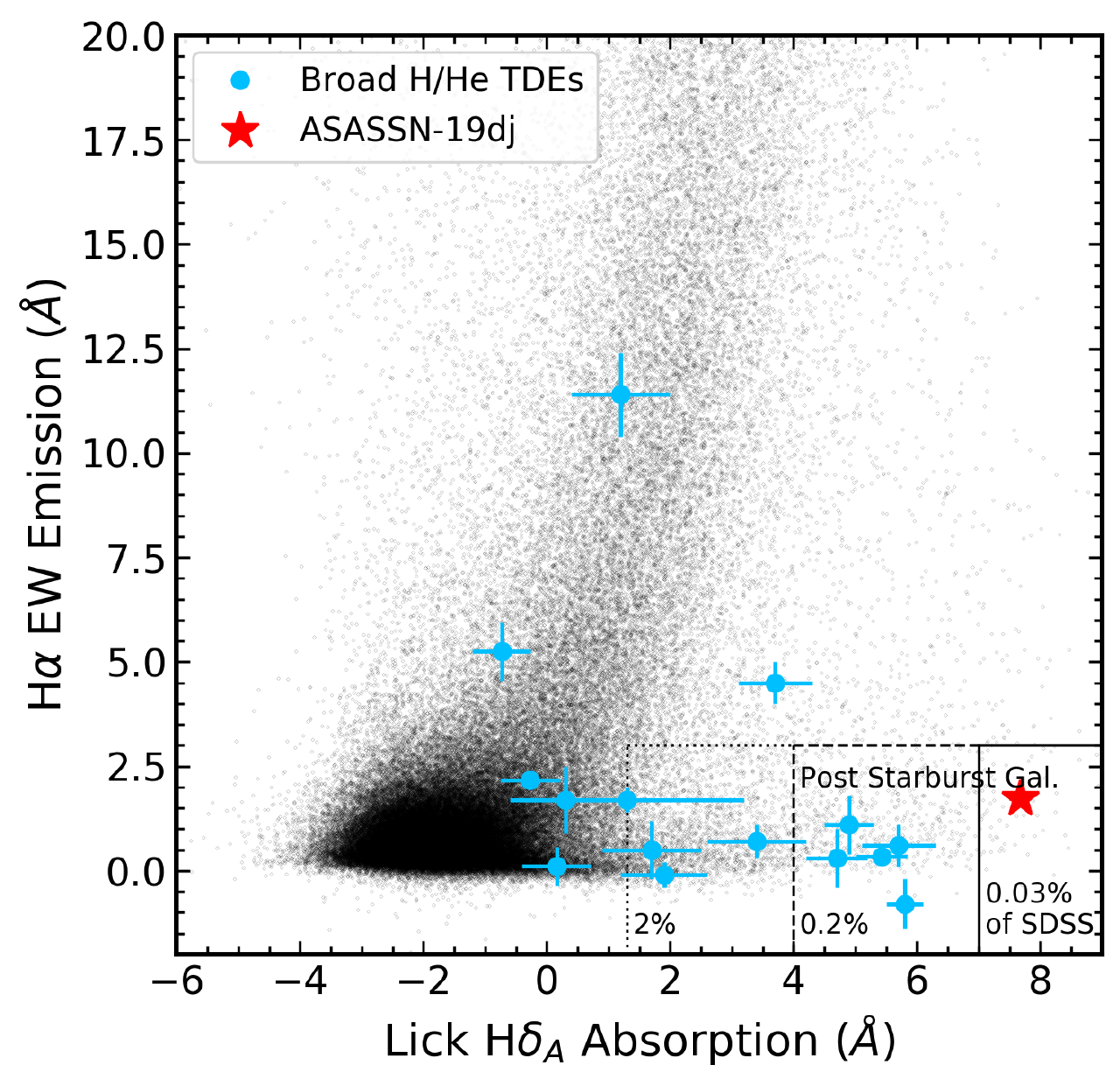}\hfill
 \includegraphics[width=.48\textwidth]{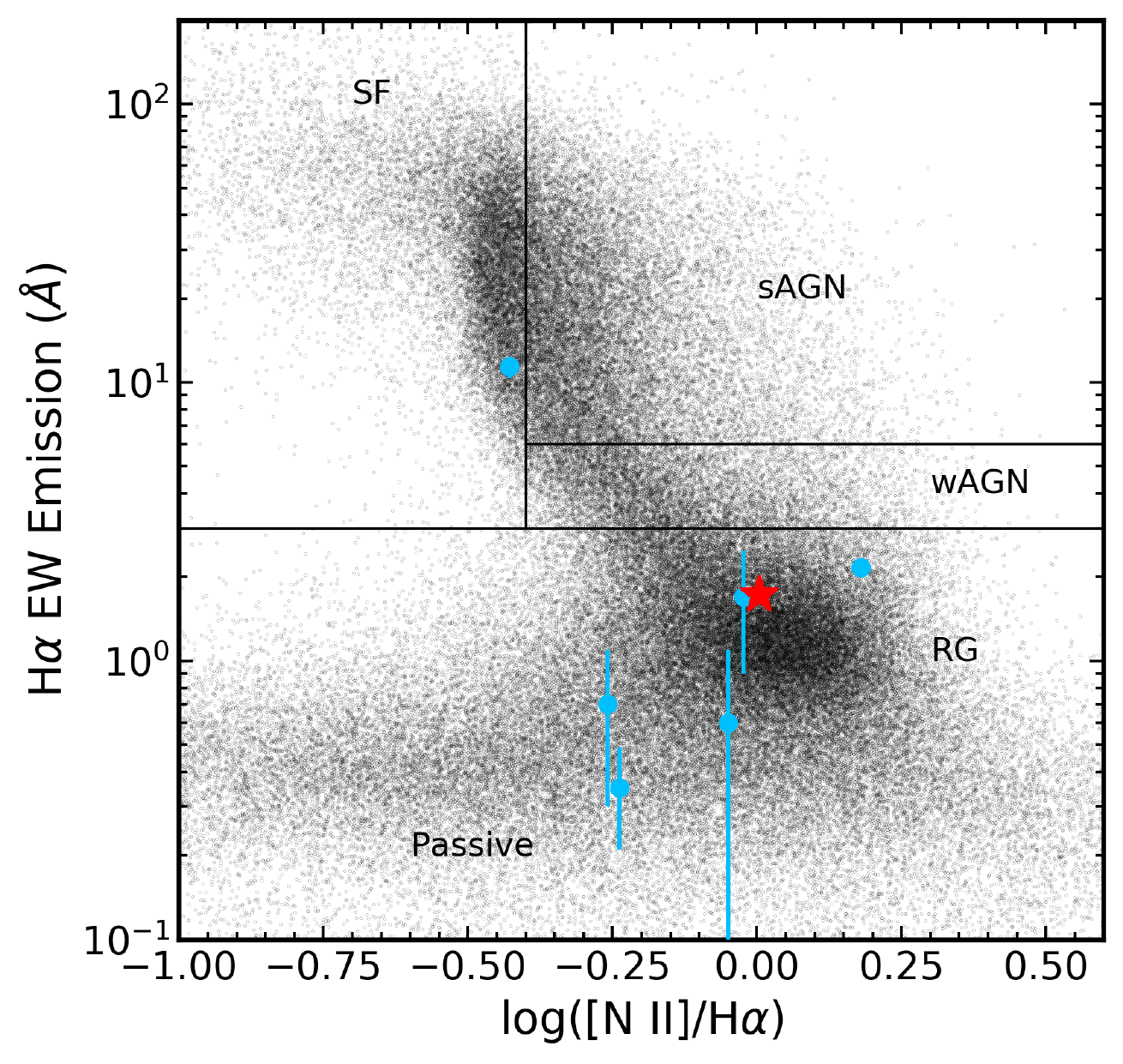} \\
 \includegraphics[width=.48\textwidth]{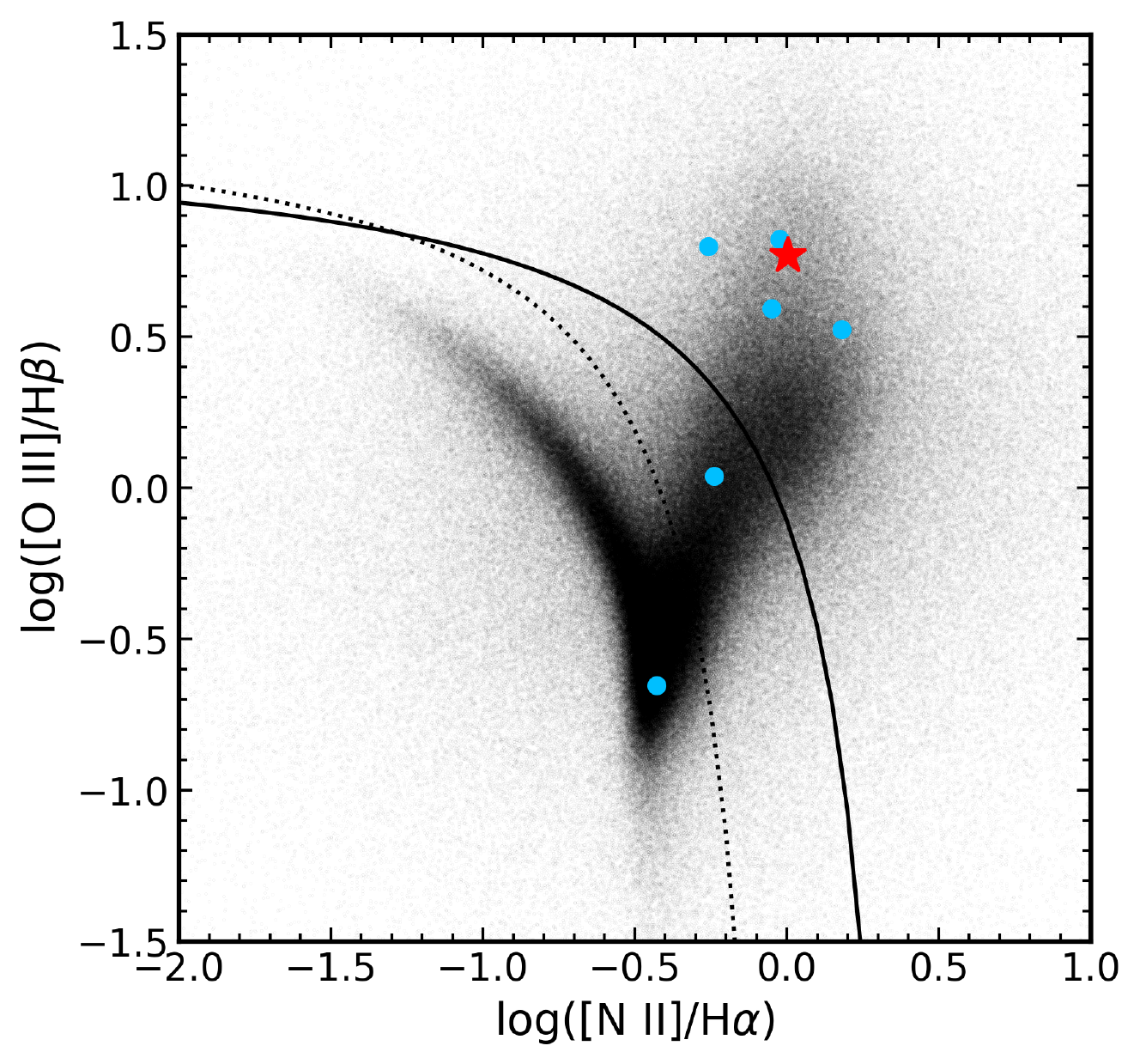}\hfill
 \includegraphics[width=.48\textwidth]{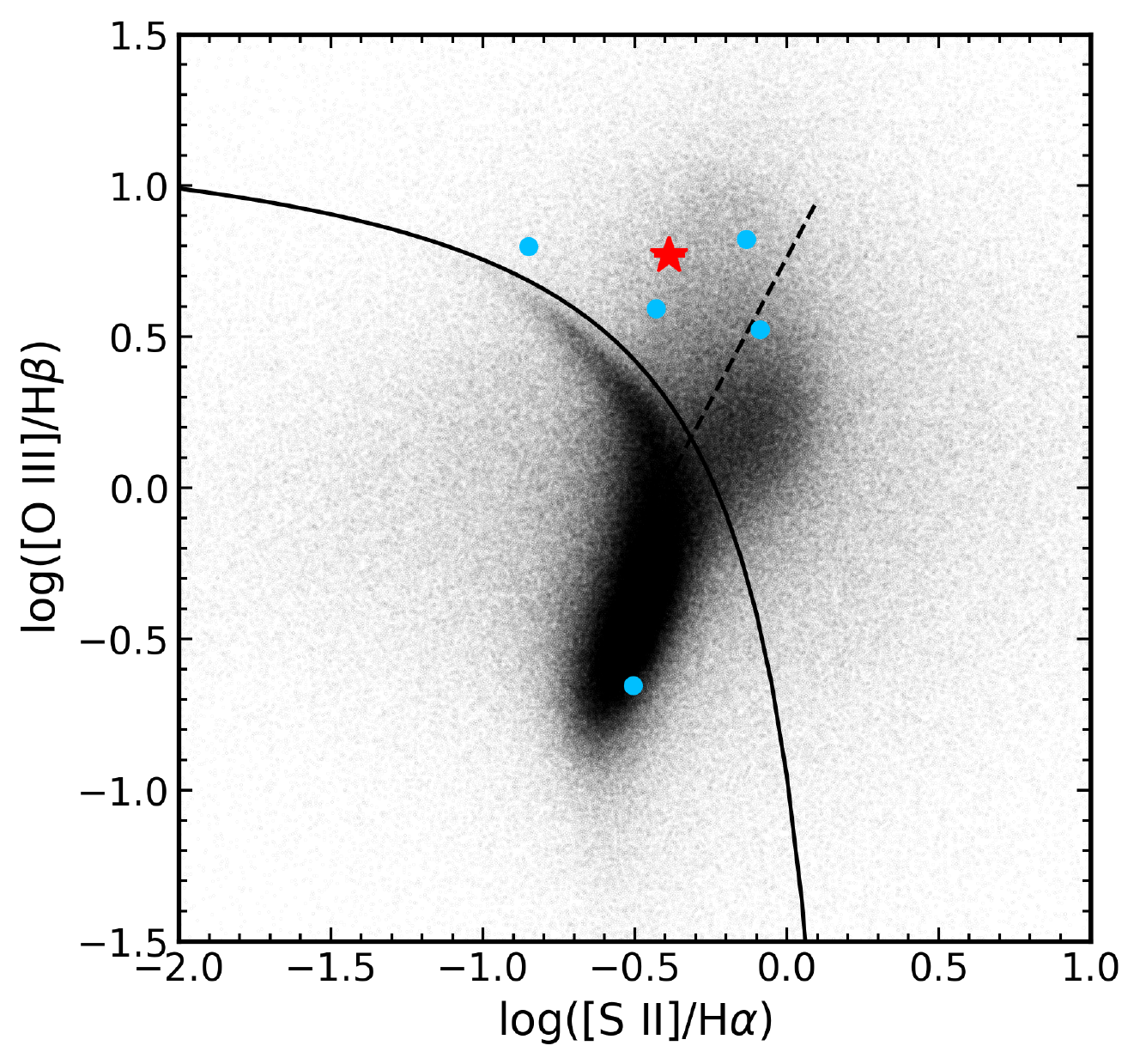}
 \caption{\textit{Upper Left Panel}: H$\alpha$ emission line equivalent width (EW), which traces current star formation, as compared to the Lick H$\delta_A$ absorption index, which traces star formation in the past Gyr. The host galaxy KUG 0810+227 is shown as a red star, with other TDE hosts shown as blue circles. KUG 0810+227 is similar to some extreme post-starbursts galaxies seen in SDSS. The error bars on KUG 0810+227 are roughly the size of the symbol. \textit{Upper Right Panel}: H$\alpha$ emission line equivalent width (W$_{H\alpha}$), as compared to log$_{10}$([\ion{N}{ii}] / \halpha), otherwise known as the WHAN diagram \citep{cidfernandes11}. Lines separating star-forming galaxies (SF), strong AGN (sAGN), weak AGN (wAGN), and passive and ``retired galaxies'' (RG) are shown \citep{cidfernandes11}. \textit{Lower Left Panel}: log$_{10}$([\ion{O}{iii}] / \hbeta) vs. log$_{10}$([\ion{N}{ii}] / \halpha) diagram \citep{baldwin81, veilleux87}. The solid line is the theoretical line separating AGN (above right) and H II-regions (below left) from \citet{kewley01}. The dotted line is the empirical line from \citet{kauffmann03} showing the same separation. Objects between the dotted and solid lines are classified as composites. \textit{Lower Right Panel}: log$_{10}$([\ion{O}{iii}] / \hbeta) vs. log$_{10}$([\ion{S}{ii}] / \halpha) diagram \citep{veilleux87}. The solid line is the theoretical line separating AGN (above right) and H II-regions (below left) from \citet{kewley01}. The diagonal dashed line is the theoretical line separating Seyferts (above left) and LINERs (below right) from \citet{kewley06}. KUG 0810+227 appears in the AGN/Seyfert region of both diagrams. In all panels, galaxies from SDSS Data Release 8 \citep{eisenstein11} are shown in black.}
 \label{fig:ew_bpt}
\end{figure*}

\begin{table}
\centering
 \caption{Archival Host Galaxy Photometry}
 \label{tab:arch_phot}
 \begin{tabular}{ccc}
  \hline
  Filter & Magnitude & Magnitude Uncertainty\\
  \hline
  $NUV$ & 18.71 & 0.05 \\
  $u$ & 16.80 & 0.10 \\
  $g$ & 15.12 & 0.04 \\
  $r$ & 14.59 & 0.03 \\
  $i$ & 14.35 & 0.03 \\
  $z$ & 14.13 & 0.03 \\
  $J$ & 13.94 & 0.04 \\
  $H$ & 13.99 & 0.09 \\
  $K_S$ & 14.34 & 0.05 \\
  $W1$ & 15.07 & 0.03 \\
  $W2$ & 15.70 & 0.03 \\
  \hline
 \end{tabular}\\
\begin{flushleft}Archival magnitudes of the host galaxy KUG 0810+227. $ugriz$ and $JHK_S$ magnitudes are 15\farcs{0} aperture magnitudes measured from SDSS and 2MASS images, respectively. The $NUV$ magnitude is taken from the GALEX AIS and the $W1$ and $W2$ magnitudes are taken from the WISE AllWISE catalog. All magnitudes are presented in the AB system. \end{flushleft}
\end{table}

In order to constrain the possibility of the host galaxy being an AGN, we analysed a range of archival data for KUG 0810+227. Using ROSAT All-Sky Survey (RASS) data, we find no emission from the host galaxy at a 3$\sigma$ upper-limit of $3.4 \times 10^{-2}$ counts s$^{-1}$. Assuming an AGN with a photon index of $\Gamma = 1.75$ \citep{ricci17} and a Galactic column density of $N_{H}=4.16\times10^{20}$ cm$^{-2}$ along the line of sight \citep{HI4PI16}, this corresponds to an unabsorbed flux of $1.2 \times 10^{-12} \text{ erg } \text{cm}^{-2} \text{ s}^{-1}$ in the 0.3 - 10 keV band. At the distance of KUG 0810+227, this yields an X-ray luminosity of $1.4 \times 10^{42} \text{ erg } \text{ s}^{-1}$. This limit rules out strong AGN activity, but does not rule out the presence of a weak or low luminosity AGN \citep[LLAGN; ][]{tozzi06, marchesi16, liu17, ricci17}. The mid-infrared (MIR) colour of the host ($W1-W2$) = $0.62 \pm 0.04$ mag again suggests that KUG 0810+227 does not harbour a strong AGN \citep[e.g.][]{assef13}, but still does not rule out the presence of a LLAGN where the host light dominates over light from the AGN. When fitting a flat line to the WISE $W1$ and $W2$ light curves, we obtain reduced $\chi^2$ values of 2.1 and 3.0 respectively, indicating a low level of variability consistent with a LLAGN.

We fit stellar population synthesis models to the archival photometry of KUG 0810+227 (shown in Table~\ref{tab:arch_phot}) using the Fitting and Assessment of Synthetic Templates \citep[\textsc{Fast};][]{kriek09} to obtain an SED of the host. Our fit assumes a \citet{cardelli88} extinction law with $\text{R}_{\text{V}} = 3.1$ and Galactic extinction of $\text{A}_{\text{V}} = 0.122$ mag \citep{schlafly11}, a Salpeter IMF \citep{salpeter55}, an exponentially declining star-formation rate, and the \citet{bruzual03} stellar population models. Based on the \textsc{Fast} fit, KUG 0810+227 has a stellar mass of M$_* = 9.3^{+3.0}_{-1.2} \times 10^9$~\msun, an age of $1.4^{+0.6}_{-0.5}$ Gyr, and an upper limit on the star formation rate of SFR $\leq6.9 \times 10^{-2}$ M$_{\odot}$ yr$^{-1}$. The best fit age is slightly higher than the stellar ages of other TDE host galaxies \citep[$\sim 0.5$ Gyr; ][]{french17}. Using the sample of \citet{mendel14} to compute a scaling relation between stellar mass and bulge mass, we estimate a bulge mass of $\sim 10^{9.7}$~\msun. We then use the M$_B$ - M$_{BH}$ relation of \citet{mcconnell13} to estimate a black hole mass of $\sim 10^{7.1}$~\msun, roughly a factor of three higher than that estimated by \citet{vanvelzen19}, although these methods use different data and scaling relations.

Our photometric follow-up campaign includes several filters for which archival imaging data are not available, including the \swift UVOT and $BV$ filters. In order to estimate the host flux in these filters for host flux subtraction, we convolved the host SED from \textsc{Fast} with the filter response curve for each filter to obtain 15\farcs{0} fluxes. To estimate uncertainties on the estimated host galaxy fluxes, we perturbed the archival host fluxes assuming Gaussian errors and ran 1000 different \textsc{Fast} iterations. These synthetic fluxes were then used to subtract the host flux in our non-survey follow-up data.

The upper left panel of Figure \ref{fig:ew_bpt} compares the H$\alpha$ emission line equivalent width to the Lick H$\delta_A$ absorption index, which compares current and past star formation to identify post-starburst galaxies. The upper right panel of \ref{fig:ew_bpt} shows the H$\alpha$ emission equivalent width as compared to the log$_{10}$([\ion{N}{ii}]$/$\halpha) line ratio to separate ionization mechanisms, particularly those associated with LINER-like (Low-Ionization Nuclear Emission-line Region) emission line ratios. The bottom two panels of Figure \ref{fig:ew_bpt} show log$_{10}$([\ion{O}{iii}]$/$\hbeta) vs. log$_{10}$([\ion{N}{ii}]$/$\halpha) and log$_{10}$([\ion{O}{iii}]$/$\hbeta) vs. log$_{10}$([\ion{S}{ii}]$/$\halpha) line ratios to characterise the activity of the host galaxies of TDEs. The background points in these figures are taken from the MPA-JHU catalog \citep{brinchmann04}, which calculated the spectral properties of galaxies in SDSS DR8 \citep{eisenstein11}.

We obtained the archival SDSS \citep{york00} spectrum of KUG 0810+227. This spectrum shows [\ion{N}{ii}] $\lambda$6584, [\ion{S}{ii}] $\lambda\lambda$6717, 6731, [\ion{O}{i}] $\lambda$6300, and [\ion{O}{iii}] $\lambda\lambda$4959, 5007 in emission with weak \halpha emission and \hbeta in absorption. We use the fits from the MPA-JHU catalog \citep{brinchmann04}, which model and subtract the stellar component for robust emission line fits in our further analysis of KUG 0810+227. The line ratios log$_{10}$([\ion{O}{iii}]$/$\hbeta) = 0.767, log$_{10}$([\ion{N}{ii}]$/$\halpha) = 0.004, log$_{10}$([\ion{S}{ii}]$/$\halpha) = $-0.386$, and log$_{10}$([\ion{O}{i}]$/$\halpha) = $-1.18$ place this galaxy in the AGN/Seyfert regions of the \citet{baldwin81} and \citet{ veilleux87} diagrams ([\ion{O}{iii}] $/$\hbeta vs.\ [\ion{N}{ii}]$/$\halpha, [\ion{S}{ii}]$/$\halpha, and [\ion{O}{i}]$/$\halpha). However, the WHAN diagram of Figure \ref{fig:ew_bpt} places KUG 0810+227 in the ``retired galaxies'' (RG) region, where galaxies have ceased actively forming stars and are predominantly ionised by hot, evolved, lower mass stars such as post-AGB stars \citep{cidfernandes11}. Other TDE hosts also tend to populate the RG region of this diagnostic diagram. Additionally, the observed line ratios of KUG 0810+227 can also be produced by large-scale shocks \citep[e.g.][]{rich11, rich15}. Thus, while KUG 0810+227 may harbor a LLAGN, it is also possible that other processes are at play.

From Figure \ref{fig:ew_bpt}, we see that KUG 0810+227 is a post-starburst galaxy. The archival SDSS spectrum displays weak H$\alpha$ emission and extremely strong H$\delta$ absorption, with a Lick H$\delta_{A}$ index of $7.67 \pm 0.17$ \AA \ as measured by \citet{brinchmann04}, confirming this classification. This is consistent with the tendency for TDEs to be found in post-starburst, or ``quiescent Balmer-strong'', host galaxies \citep[e.g.][]{arcavi14, french16, law-smith17}. Additionally, the host of ASASSN-19dj is similar to many other TDE hosts in terms of its star formation history. Compared to other TDE hosts, KUG 0810+227 is also similar in its line ratios. The possibility that KUG 0810+227 hosts an LLAGN is in line with the fact that hosts of other TDE such as ASASSN-14ae \citep{holoien14b}, ASASSN-14li \citep{holoien16a, french20}, and ASASSN-19bt \citep{holoien19c} show evidence for weak AGN activity.

The fact that KUG 0810+227 is a RG is in line with several studies on the ionization processes in post starburst galaxies. \citet{depropris14} compiled a sample of ten post-starburst galaxies with Hubble Space Telescope (HST) imaging, optical spectra, X-ray, far-infrared, and radio data. They found no evidence of AGN down to an Eddington ratio of 0.1\% in these galaxies. Similarly, \citet{french18} found that many post-starburst galaxies have LINER-like line ratios and that most are in the RG region of the WHAN diagram. The TDE hosts in particular have lower H$\alpha$ EW, placing them solidly in the RG region \citep{french17}. However, \citet{prieto16} suggests that the host galaxy of the TDE ASASSN-14li, which we note is a RG in the WHAN diagram, may host an AGN.

\begin{figure*}
\centering
 \includegraphics[width=1.0\textwidth]{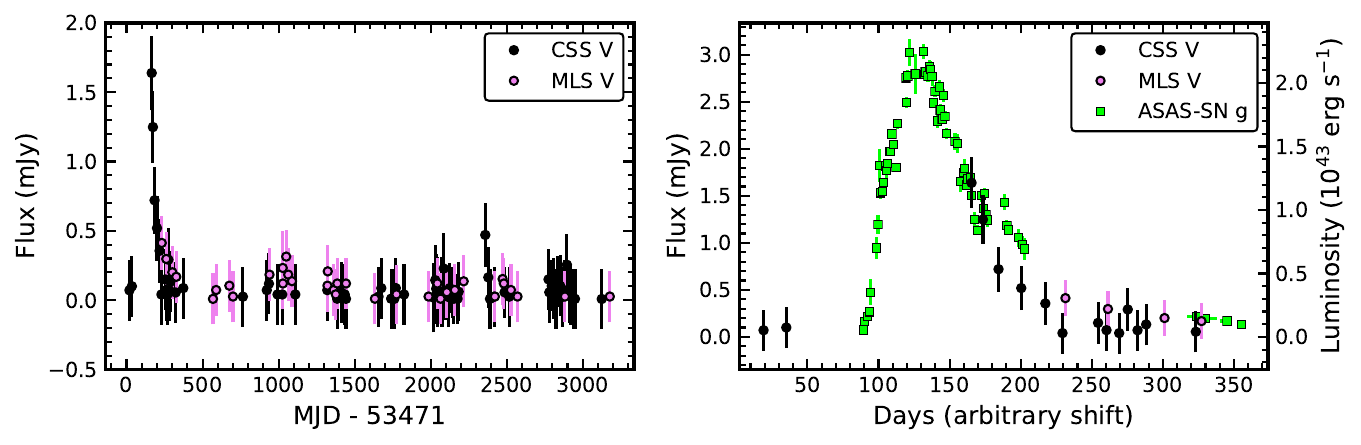}\hfill
 \caption{\textit{Left Panel}: The $V$-band CRTS data (CSS in black and MLS in violet) light curves. \textit{Right Panel}: Comparison of the flare seen at MJD $\sim$ 53680 by CRTS and the ASAS-SN $g$-band data (shown in green) of the TDE ASASSN-19dj with an arbitrary shift in days between the two flares so that they are roughly aligned. The two light curves are offset by their best-fit zero-points and have been corrected for Galactic extinction.}
 \label{fig:crts}
\end{figure*}

Archival Catalina Real-Time Transient Survey \citep[CRTS;][]{drake09} data indicates that KUG 0810+227 experienced an outburst at MJD $\sim$ 53640 (September 2005), roughly 14.5 years prior to ASASSN-19dj. We obtained photometric data for this flare from both the Catalina Sky Survey (CSS) 0.7-m and the Mount Lemmon Survey (MLS) 1.5-m telescopes. First, we fit a flat line to the CRTS data between MJD = 54592 and MJD = 55919, outside of the flare to obtain a flux zero point. The zero-point-subtracted light curve and a comparison of the flare to ASASSN-19dj are shown in Figure \ref{fig:crts}. The reduced chi-squared of the non-outburst parts of the CRTS light curve as compared to the zero-point fit is 1.38, indicating that the $V$-band light curve is non-variable. The flare in CRTS appears to be of a similar magnitude to ASASSN-19dj, though the peak of the CRTS flare may have occurred in the seasonal gap. The FWHM (full width at half maximum) of the  ASASSN-19dj flare is $\sim80$ days in the ASAS-SN $g$-band data, while the FWHM of the CRTS flare is $\la 140$ days, although we do not see the full rise or peak of this flare. Finally, the light curve of ASASSN-19dj appears to decline slower than the CRTS flare as indicated by the right panel of Figure \ref{fig:crts}. 

In addition to our search of CRTS data, we searched archival Pan-STARRS data for previous outbursts. In this search, we found no clear detections prior to ASASSN-19dj. This places strong constraints on any potential flaring activity of the host galaxy, with two large ($\sim$15-16 mag) flares and otherwise no significant optical variability over the course of roughly 15 years.

The photometry from CRTS uses SExtractor \citep[][]{bertin96}, which includes flux from the entire host galaxy. Thus it is not possible to determine the nature of the transient from the CRTS photometry alone. In an effort to link this flare to a known transient, we searched multiple databases of known supernovae such as the Open Supernova Project \citep{Guillochon17} and the Central Bureau for Astronomical Telegrams (CBAT\footnote{\url{http://www.cbat.eps.harvard.edu/lists/Supernovae.html}}) but found no reported SNe or other flares at the time. Nevertheless, even if the source is consistent with the nucleus, it is difficult to constrain the nature of the earlier transient without spectroscopic or multi-band photometric observations during that epoch.

\subsection{ASAS-SN Light Curve}
ASAS-SN is a fully automated transient survey which consists of 20 telescopes on 5 robotic mounts. Each telescope is a 14-cm aperture Nikon telephoto lens with 8\farcs{0} pixels, and each unit consists of 4 telescopes on a common mount. Single ASAS-SN units are located at Haleakal\=a Observatory, McDonald Observatory, and the South African Astrophysical Observatory (SAAO), and two are located at Cerro Tololo Inter-American Observatory (CTIO). With our current network, ASAS-SN monitors the visible sky with a cadence of $\sim 20$ hours to a depth of $g \sim 18.5$ mag.  

Since ASASSN-19dj is near the equator, it was observed from both hemispheres and by all 5 units. Additionally, ASASSN-19dj lies in a designed 0.5 degree field overlap region, giving us roughly twice the cadence. Thus we observed ASASSN-19dj with 10 of the 20 ASAS-SN cameras currently deployed. However, the filters for two of the ASAS-SN units, Brutus and Cassius, had been changed from $V$ to $g$ shortly before ASASSN-19dj was discovered. Thus, there were not enough $g$-band images taken before the TDE to construct a subtraction reference image for those cameras. Because of this we had to modify the normal ASAS-SN processing pipeline to extract the light curve. 

Images were reduced using the automated ASAS-SN pipeline but we performed image subtraction separately. We used the ISIS image subtraction package \citep{Alard1998, Alard2000} with the same parameters as the ASAS-SN pipeline, but images from all cameras for a given pointing were first interpolated onto a common grid. We then built a reference image using good images from multiple cameras observed well before the rise of ASASSN-19dj. This common reference image was used to analyze all the data.

We then used the IRAF {\tt apphot} package with a 2-pixel radius (approximately 16\farcs{0}) aperture to perform aperture photometry on each subtracted image, generating a differential light curve. The photometry was calibrated using the AAVSO Photometric All-Sky Survey \citep{Henden2015}. We visually inspected each of the 1155 exposures taken after 2018 February 5 analyzed in this work for clouds or flat-fielding issues and disregarded any where issues were seen. We also discarded images with a FWHM of 1.67 pixels or greater.

To increase the signal-to-noise ratio (S/N) we stacked our photometric measurements. For exposures with TDE emission, the photometric measurements were stacked in 12 hour bins. Prior to our first detection, measurements within 75 hours of each other were stacked to get deeper upper limits. After the 2019 seasonal gap, measurements within 10 days of each other were stacked to sample the decline of the TDE. 

\subsection{ATLAS light curve}
\label{sec:ATLAS_LC}
The ATLAS survey is designed primarily to detect small (10--140 m) asteroids that may collide with Earth \citep{tonry18}. ATLAS uses two 0.5m f/2 Wright Schmidt telescopes on Haleakal\=a and at the Mauna Loa Observatory. For normal operation, the telescopes obtain four 30-second exposures of 200--250 fields per night. This allows the telescopes to cover roughly a quarter of the sky visible from Hawaii each night, ideal for transient detection \citep{smith20}. ATLAS uses two broad-band filters, the `cyan' ($c$) filter from 420--650 nm and the `orange' ($o$) filter covering the 560--820 nm range \citep{tonry18}.

Each ATLAS image is processed by a pipeline that performs flat-field corrections in addition to astrometric and photometric calibrations. Reference images of the host galaxy were created by stacking multiple images taken under excellent conditions before MJD = 58251 and this reference was then subtracted from each science image of ASASSN-19dj in order to isolate the flux from the transient. We performed forced photometry on the subtracted ATLAS images of ASASSN-19dj as described in \citet{tonry18}. We combined the four intra-night photometric observations using a weighted average to get a single flux measurement. The ATLAS $o$-band photometry and 3-sigma limits are presented in Table \ref{tab:phot} and are shown in Figure~\ref{fig:opt_lc}. We do not plot the $c$-band photometry in Figure~\ref{fig:opt_lc} as there were few $c$-band observations in the rise to peak and near peak due to weather and the design of the ATLAS survey, but as they provide useful early limits and detections, we present them in Table \ref{tab:phot}.

\subsection{ZTF Light Curves}
The ZTF survey uses the Samuel Oschin 48" Schmidt telescope at Palomar Observatory and a camera with a 47 square degree field of view that reaches as deep as 20.5 $r$-band mag in a 30 second exposure. Alerts for transient detection from ZTF are created from the final difference images \citep{patterson19}. These alerts are distributed to brokers including Lasair \citep{smith19} through the University of Washington Kafka system. For ASASSN-19dj, we obtained ZTF $g$- and $r$-band light curves from the Lasair broker\footnote{\url{https://lasair.roe.ac.uk/}}. Lasair uses ZTF difference imaging photometry so the host flux is subtracted. The ZTF magnitudes presented in this paper are calculated using PSF photometry. Similar to the ATLAS data, we combined the intra-night photometric observations using a weighted average to get a single flux measurement.

\subsection{Additional Ground-Based Photometry}
We also obtained photometric follow-up observations from several ground-based observatories. We used the Las Cumbres Observatory \citep{brown13} 1-m telescopes located at CTIO, SAAO, McDonald Observatory, and Siding Spring Observatory for $BVgri$ observations, and the Swope 1-m telescope at Las Campanas Observatory for $uBVgri$ observations. After applying flat-field corrections, we solved astrometry in each image using astrometry.net \citep{barron08,lang10}. 

We aligned the $ugri$ data to the archival SDSS image in the corresponding filter for each follow-up image using the Python {\tt reproject} package, which uses the WCS information of two images to project one image onto the other. We then subtracted the SDSS template images from each follow-up image using \textsc{Hotpants}\footnote{\url{http://www.astro.washington.edu/users/becker/v2.0/ hotpants.html}}\citep{becker15}, an implementation of the \citet{alard00} image subtraction algorithm, and used the IRAF {\tt apphot} package to measure 5\farcs{0} aperture magnitudes of the transient. For the $BV$ data, we did not have archival images available to use as subtraction template images. Instead, we used {\tt apphot} to measure 15\farcs{0} aperture magnitudes of the host $+$ transient, and subtracted the 15\farcs{0} host flux synthesised from our \textsc{FAST} fit in the appropriate filter to isolate the transient flux. For all filters, we used SDSS stars in the field to calibrate our photometry, using the corrections from \citet{lupton05} to calibrate the $B$ and $V$ band magnitudes with the $ugriz$ data.

We measured the centroid position of the transient in a host-subtracted Las Cumbres Observatory $g$-band image taken near peak using the IRAF {\tt imcentroid} package. This yielded a position of $(\alpha,\delta)=($08:13:16.96$,+$22:38:54.00$)$. We also used the archival SDSS $g$-band image to measure the position of the nucleus of KUG 0810+227, finding $(\alpha,\delta)=($08:13:16.95$,+$22:38:53.89$)$. This gives an angular offset of 0\farcs{21}$\pm$0\farcs{12}, where the uncertainty is due to uncertainty in the centroid positions of the TDE and host nucleus. We also measured the centroid positions of several stars in both the follow-up and SDSS host images, finding that the stars had an average random offset of 0\farcs{19}. Combining these sources of uncertainty, the transient position is thus offset 0\farcs{21}$\pm$0\farcs{24} from the position of the host nucleus, corresponding to a physical distance of $99.2\pm112.5$~pc.

\begin{table}
\centering
 \caption{Host-Subtracted Photometry of ASASSN-19dj}
 \label{tab:phot}
 \begin{tabular}{ccccc}
  \hline
  MJD & Filter & Magnitude & Uncertainty & Telescope\\
  \hline
  58537.07 & $i$ & 16.36 & 0.01 & LCOGT-1m \\
  58539.86 & $i$ & 16.24 & 0.01 & LCOGT-1m \\
  58541.81 & $i$ & 16.12 & 0.01 & LCOGT-1m \\
  \ldots & \ldots & \ldots & \ldots & \ldots \\
  58850.21 & $UVW2$ & 17.68 & 0.07 & \swift \\
  58906.65 & $UVW2$ & 17.74 & 0.11 & \swift \\
  58911.89 & $UVW2$ & 18.00 & 0.18 & \swift \\
  \hline
 \end{tabular}\\
\begin{flushleft}Host-subtracted magnitudes and 3$\sigma$ upper limits for all follow-up photometry. A range of MJD in the first column indicates the beginning and end of the range over which data were stacked to increase S/N. All magnitudes are corrected for Galactic extinction and presented in the AB system. The last column reports the source of the data for each epoch. The \swift $B$ data do not include the shift applied in Figure \ref{fig:opt_lc}. Only a small section of the table is displayed here. The full table can be found online as an ancillary file.\end{flushleft}
\end{table}

\subsection{\textit{Swift} Observations}
Fourty-four total Neil Gehrels Swift Gamma-ray Burst Mission (\textit{Swift}; \citealt{gehrels04}) target-of-opportunity (ToO) observations were carried out between 2019 March 2 and 2020 March 3 (Swift target ID 11186 (as AT2019azh; PIs: Gezari, Arcavi, and Wevers), and Swift target ID 12174 (as ASASSN-19dj; PI: Hinkle). These observations used the UltraViolet and Optical Telescope (UVOT; \citealt{roming05}) and X-ray Telescope (XRT; \citealt{burrows05}) to study the multiwavelength properties of the TDE.

\subsubsection{UVOT Observations}
For a majority of the observation epochs, \swift{} observed ASASSN-19dj with all six UVOT filters \citep{poole08}: $V$ (5468 \AA), $B$ (4392 \AA), $U$ (3465 \AA), $UVW1$ (2600 \AA), $UVM2$ (2246 \AA), and $UVW2$ (1928 \AA). Each epoch of UVOT data includes 2 observations in each filter, which we combined into one image for each filter using the HEASoft {\tt uvotimsum} package. We then used the {\tt uvotsource} package to extract source counts using a 15\farcs{0} radius region centered on the position of the TDE and background counts using a source-free region with radius of $\sim$40\farcs{0}. We converted the UVOT count rates into fluxes and magnitudes using the most recent calibrations \citep{poole08, breeveld10}.

Because the UVOT uses unique $B$ and $V$ filters, we used publicly available colour corrections\footnote{\url{https://heasarc.gsfc.nasa.gov/docs/heasarc/caldb/swift/docs/uvot/uvot_caldb_coltrans_02b.pdf}} to convert the UVOT $BV$ data to the Johnson-Cousins system. We then corrected the UVOT photometry for Galactic extinction and removed host contamination by subtracting the corresponding 15\farcs{0} host flux in each filter, as we did with the ground-based $BV$ data.

\begin{figure*}
\centering
 \includegraphics[width=0.99\textwidth]{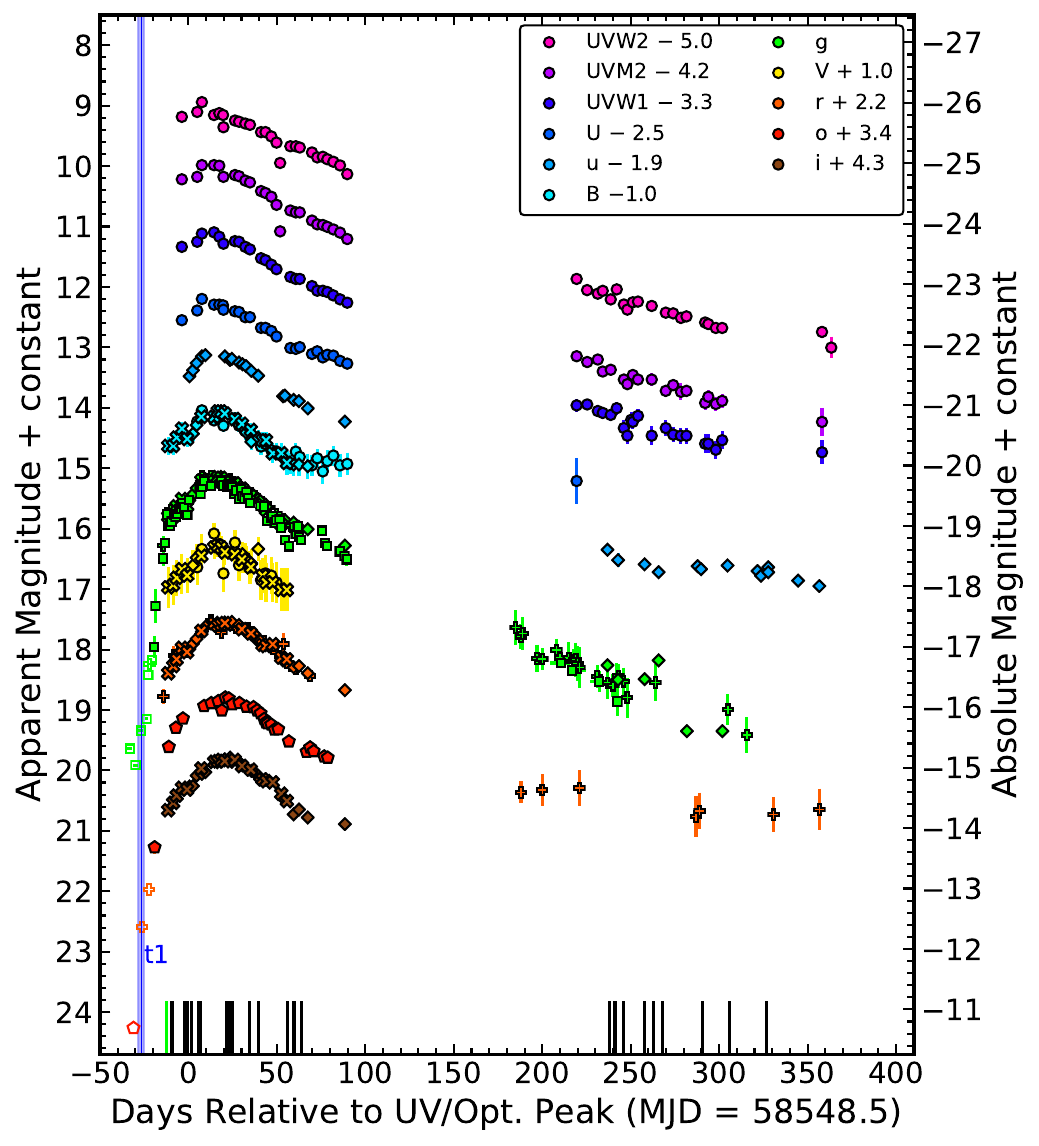}
 \caption{Host-subtracted UV and optical light curves of ASASSN-19dj, showing ASAS-SN ($g$, squares), Swift (UV+$UBV$, circles), ATLAS ($o$, pentagons), ZTF ($gr$, pluses), Swope ($uBVgri$, diamonds), and Las Cumbres Observatory 1-m telescopes ($BVgri$, x-shapes) photometry spanning from roughly 20 days prior to peak (MJD = 58548.5) to roughly 390 days after in observer-frame days. Horizontal error bars on the ASAS-SN $g$-band data indicate the date range of observations stacked to obtain deeper limits and higher S/N detections, although they are small and difficult to see. Open symbols indicate 3$\sigma$ upper limits when the TDE was not detected. The green bar on the x-axis marks the epoch of ASAS-SN discovery. Black bars along the x-axis show epochs of spectroscopic follow-up. The blue line is the estimated time of first light (see \S \ref{lc}) with the shading corresponding to the uncertainty. All data are corrected for Galactic extinction and shown in the AB system.}
 \label{fig:opt_lc}
\end{figure*}

Figure \ref{fig:opt_lc} shows the extinction-corrected, host-subtracted light curves of ASASSN-19dj. The photometry spans from the shortest $UVW2$ (1928 \AA) band of Swift to $i$-band ($\sim$ 7609 \AA) from Swope and LCOGT and includes the data ranging from 21 days prior to peak to 392 days after peak. The corrected \swift $B$ data were inconsistent with the ground-based $B$ data, so we shifted the \swift $B$ data in Figure \ref{fig:opt_lc}. To do this, we computed the median offset between the LCOGT and Swope $B$-band data from the \swift $B$ data, which was found to be 0.27 mag. All the UV and optical photometry shown in Figure~\ref{fig:opt_lc}, in addition to limits not shown in this figure, is presented in Table~\ref{tab:phot}.

\subsubsection{XRT Observations}
In addition to the \textit{Swift} UVOT observations, we also obtained simultaneous \textit{Swift} X-Ray Telescope (XRT) photon-counting observations. All observations were reprocessed from level one XRT data using the \textit{Swift} \textsc{xrtpipeline} version 0.13.2, producing cleaned event files and exposure maps. Standard filter and screening criteria\footnote{\url{http://swift.gsfc.nasa.gov/analysis/xrt_swguide_v1_2.pdf}} were used, as well as the most up-to date calibration files.

To extract both background-subtracted count rates and spectra, we used a source region with a radius of 50'' centered on the position of ASASSN-19dj and a source free background region centered at ($\alpha$,$\delta$)=($08^{h}13^{m}07.93^{s},+22^{\circ}35'15.36''$) with a radius of 150''.0. The reported count rates are aperture corrected where a 50'' source radius contains $\sim90\%$ of the source counts at 1.5\,keV, assuming an on-axis pointing \citep{moretti04}. To increase the S/N of our observations, we combined our individual \textit{Swift} observations into six time bins using \textsc{XSELECT} version 2.4g, allowing us to extract spectra with $\gtrsim200-300$ background subtracted counts. 

To extract spectra from our merged observations, we used the task \textsc{xrtproducts} version 0.4.2 and the regions defined above to extract both source and background spectra. To extract ancillary response files (ARF), we first merged the corresponding individual exposure maps that were generated by \textsc{xrtpipeline} using \textsc{XIMAGE} version 4.5.1 and then used the task \textsc{xrtmkarf}. We used the ready-made response matrix files (RMFs) that are available with the \textit{Swift} calibration files. Each spectrum was grouped to have a minimum of 10 counts per energy bin using the \textsc{FTOOLS} command \textsc{grppha}.

\subsection{\textit{NICER} Observations}
After the 2019 seasonal gap, follow-up \textit{Swift} XRT  observations found that the X-ray flux of ASASSN-19dj had increased by an order of magnitude compared to the flux approximately 100 days earlier (see \S \ref{sec:xray}). ToO observations of ASASSN-19dj were then obtained using the Neutron star Interior Composition ExploreR (NICER: \citealt{gendreau12}), which is an external payload on the International Space station that has a large effective area over the 0.2-12.0 keV energy band and provides fast X-ray timing and spectroscopic observations of sources. In total, 80 observations were taken between 2019 October 23 and 2020 March 12 (Observation IDs: 2200920101--2200920176, 3200920101--3200920105, PI:Pasham/Gendreau, \citealt{2019ATel13221....1P}), totaling 169 ks of cumulative exposure. 

The data were reduced using \textsc{NICERDAS} version 6a, \textsc{HEASOFT} version 6.26.1. Standard filtering criteria were applied using the \textsc{NICERDAS} task \textsc{nicerl2}. Here the standard filter criteria includes\footnote{See \url{https://heasarc.gsfc.nasa.gov/docs/nicer/data_analysis/nicer_analysis_guide.html} or \citep{2019ApJ...887L..25B} for more details about these criteria.}: the \textit{NICER} pointing is (ANG\_DIST) $<$0.015 degrees from the position of the source; excluding events that were acquired during passage through the South Atlantic Anomaly, or those that are obtained when Earth was 30$^{\circ}$ (40$^{\circ}$) above the dark (bright) limb (ELV and BR\_EARTH, respectively). We also removed events that are flagged as overshoot, or undershoot events (EVENT\_FLAGS=bxxxx00), and we used the so-called ``trumpet filter'' to remove events with a PI\_RATIO $>$ 1.1+120/PI, where PI is the pulse invariant amplitude of an event, as these are likely particle events \citep{2019ApJ...887L..25B}. To extract time-averaged spectra and count rates, we used \textsc{XSELECT}, and ready made ARF (nixtiaveonaxis20180601v002.arf) and RMF (nixtiref20170601v001.rmf) files that are available with the \textit{NICER} CALDB. Similar to the \textit{Swift} spectra, each spectrum was grouped with a minimum of 10 counts per energy bin. As \textit{NICER} is a non-imaging instrument, background spectra were generated using the background modeling tool \textsc{nibackgen3C50}\footnote{\url{https://heasarc.gsfc.nasa.gov/docs/nicer/tools/nicer_bkg_est_tools.html}}.

To analyse the spectra extracted from both our \textit{Swift} and \textit{NICER} observations, we used the X-ray spectral fitting package (XSPEC) version 12.10.1f \citep{arnaud96} and $\chi^{2}$ statistics. Both the \textit{Swift} and \textit{NICER} data and their analysis are further discussed in \S \ref{sec:xray}. 

\subsection{\textit{XMM-Newton} slew observations}

In addition to \swift and \textit{NICER} observations, we also searched for \textit{XMM-Newton} slew observations that overlap the position of ASASSN-19dj. These slew observations are taken using the PN detector of \textit{XMM-Newton} as it maneuvers between pointed observations, detecting X-ray emission down to a 0.2-10.0 keV flux limit of $\sim10^{-12}$ erg cm$^{-2}$ s$^{-1}$ \citep{2008A&A...480..611S}. We found two slew observation (ObsIDs: 9353900003 and 9363000003) coincident with the position of ASASSN-19dj. These observations were taken on 2019-04-07 and 2019-10-05, respectively, corresponding to $\sim44$ and $\sim225$ days after discovery. To analyse these observations we follow the current slew analysis thread on the \textit{XMM-Newton} Science System (SAS) data analysis threads\footnote{\url{https://www.cosmos.esa.int/web/xmm-newton/sas-thread-epic-slew-processing}}. Here we use the SAS tool command \textsc{eslewchain} and the most up to date calibration files to produce filtered event files that we use in our analysis. Similar to our \textit{XRT} analysis, we extract the number of counts using a source region with a radius of 30'' centered on the position of ASASSN-19dj and a source free background region with a radius of 150'' centered at ($\alpha$,$\delta$)=($08^{h}13^{m}31^{s}.79,+22^{\circ}37'30''.53$). A 30'' source region contains 85\% of all source photons at 1.9 keV. Due to the low exposure times of each observation, which was determined using the corresponding exposure files of each observation, no spectra could be extracted. The 0.3 - 10 keV X-ray luminosities and hardness ratios derived from the count rates for the various X-ray epochs are shown in Table \ref{tab:xray_lum_hr}.

\begin{table}
\centering
 \caption{X-ray Luminosity and Hardness Ratios of ASASSN-19dj}
 \label{tab:xray_lum_hr}
 \begin{tabular}{cccccc}
  \hline
  MJD & log Lum. & Lum. Error & HR & HR Error & Satellite\\
   & (erg s$^{-1}$) & (erg s$^{-1}$) & & & \\
  \hline
  58544.76 & 40.72 & --- & -0.07 & --- &\swift \\
  58553.45 & 41.42 & 0.23 & -0.53 & 0.28 & \swift \\
  58556.11 & 41.36 & 0.26 & -1.00 & 0.01 & \swift \\
  \ldots & \ldots & \ldots & \ldots & \ldots \\
  58922.58 & 42.06 & 0.04 & -0.70 & 0.09 & \textit{NICER} \\
  58934.02 & 42.03 & 0.06 & -0.63 & 0.11 & \textit{NICER} \\
  58940.53 & 41.96 & 0.06 & -0.40 & 0.07 & \textit{NICER} \\
  \hline
 \end{tabular}\\
\begin{flushleft}X-ray luminositites and hardness ratios with associated uncertainties. Dashed lines represent 3$\sigma$ upper limits. The hardness ratio is defined as (H$-$S)/(H+S), where we define hard counts H as the number of counts in the 2-10 keV range and soft counts S are the number of counts in the 0.3-2 keV. The last column reports the source of the data for each epoch. Only a small section of the table is displayed here. The full table can be found online as an ancillary file.\end{flushleft}
\end{table}

\subsection{Spectroscopic Observations}
In addition to the ePESSTO spectrum released on TNS, we obtained follow-up spectra of ASASSN-19dj with the Low Dispersion Survey Spectrograph (LDSS-3) on the 6.5-m Magellan Clay telescope, the Inamori-Magellan Areal Camera and Spectrograph \citep[IMACS;][]{dressler11} on the 6.5-m Magellan-Baade telescope, the Wide Field Reimaging CCD Camera (WFCCD) on the du Pont 100-inch telescope, the SuperNova Integral Field Spectrograph \citep[SNIFS;][]{lantz04} on the 88-inch University of Hawaii telescope, the Low-Resolution Imaging Spectrometer \citep[LRIS;][]{oke95} on the 10-m Keck I telescope, the Goodman High Throughput Spectrograph \citep[GHTS; ][]{clemens04} on the 4.1-m Southern Astrophysical Research Telescope (SOAR), the Kast Double spectrograph on the Lick Shane 120-inch telescope, and the Multi-Object Double Spectrographs \citep[MODS;][]{pogge10} on the Large Binocular Telescope \citep[LBT; ][]{hill06}. Three of our spectra were obtained prior to peak light and twenty-one were obtained after peak. Most of the spectra were reduced and calibrated with standard IRAF procedures, such as bias subtraction, flat-fielding, 1-D spectroscopic extraction, and wavelength calibration. The IMACS data from 2019 November 19.3 were reduced using an updated version of the routines developed by \citet{kelson14}. The flux calibration for our observations was initially done using standard star spectra obtained on the same nights as the science spectra. We then scaled and tilted our spectra to match the calibrated flux of the TDE in the optical photometry.  

\begin{figure*}
\centering
 \includegraphics[width=1.0\textwidth]{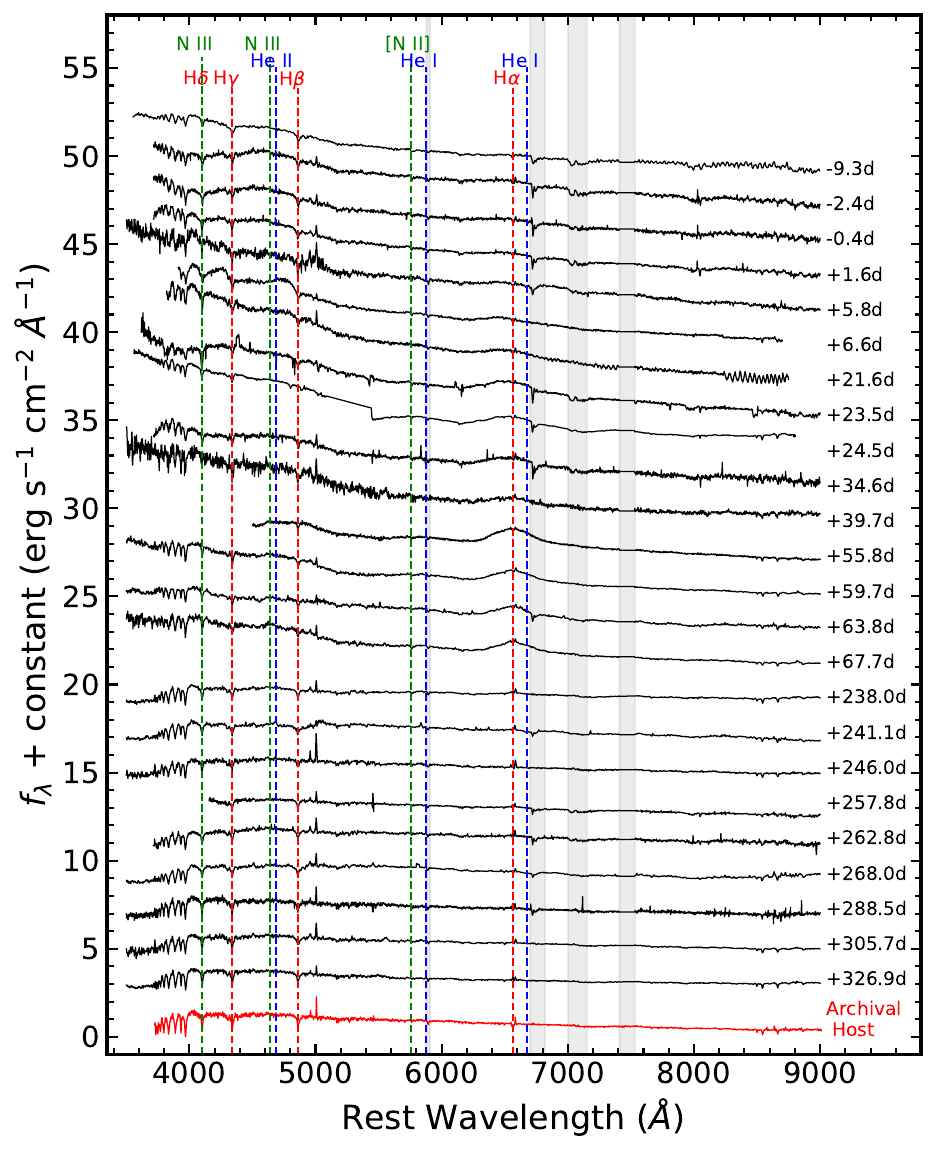}
 \caption{Optical spectroscopic evolution of ASASSN-19dj spanning from 9.3 days prior to peak UV/optical emission (top) until 326.9 days after peak (bottom). These spectra are calibrated using the photometry presented in Figure \ref{fig:opt_lc}. The vertical gray bands mark atmospheric telluric features and the strong telluric feature between $\sim$ 7400 - 7550 \AA \ has been masked. An archival host spectrum from SDSS is shown in red at the very bottom. The vertical lines mark spectral features common in TDEs, with hydrogen lines in red, helium lines in blue, and nitrogen lines in green. The straight line in the spectrum at 24.5 days after peak connects the blue and red sides of the LRIS spectrum with a large dichroic feature.}
 \label{fig:opt_spec}
\end{figure*}

All the classification and follow-up spectra for ASASSN-19dj are presented in Figure \ref{fig:opt_spec}. From top to bottom, the optical spectrum evolves from a hot, blue continuum to a host-dominated spectrum. The locations of several emission lines commonly seen in TDEs are marked with vertical dashed lines. Some of these emission lines appear, evolve, and disappear throughout the time period probed by these spectra.

\section{Analysis}\label{analysis}

\subsection{Light Curve}\label{lc}
Using Markov Chain Monte Carlo (MCMC) methods, we fit each of the epochs where there is \swift UV photometry as a blackbody to obtain the bolometric luminosity, temperature, and effective radius of ASASSN-19dj. So that our fits are relatively unconstrained, we ran each of our blackbody fits with flat temperature priors of 10000 K $\leq$ T $\leq$ 55000 K. To find the time of peak UV/optical luminosity, we fit a parabola to the light curve created by bolometrically correcting the ASAS-SN $g$-band light curve using these blackbody fits. For this fit, we excluded any upper limits. Because the curve is quite flat near peak, we fit the parabola in a narrow range between MJD = 58535.2 and MJD = 58556.2. We generated 10,000 realizations of the bolometric light curve in this date range with each magnitude perturbed by its uncertainty assuming Gaussian errors. We then fit a parabola to each of these light curves and took the median value as the peak and 16th and 84th percentiles as the uncertainties in peak time. Using this procedure, we find the time of peak bolometric luminosity to be MJD $= 58548.5^{+6.3}_{-2.6}$. From Figure \ref{fig:opt_lc}, looking from the shortest wavelength ($UVW2$) to the longest ($i$), we see that the time of peak light in each band is offset. Using a similar procedure to the bolometric light curve, but for the flux in a single photometric band, we find that the \swift $UVW2$ light curve peaks at MJD $= 58554.9^{+1.1}_{-1.5}$ and the LCOGT $i$-band light curve peaks at MJD $= 58571.9 \pm 0.1$. This offset of $\sim 17$ days is longer than the offsets seen in other TDEs such as ASASSN-18pg \citep{holoien20} and ASASSN-19bt \citep{holoien19c}. This likely occurs because the temperature of ASASSN-19dj steadily declines for roughly 25 days before increasing for 20 days and gradually leveling off for the next $\sim$ 230 days (see \S \ref{sed}), in contrast to the TDEs that exhibit relatively constant temperature near peak \citep[e.g.][]{hinkle20a, vanvelzen20}. 

\begin{figure}
\centering
 \includegraphics[width=.48\textwidth]{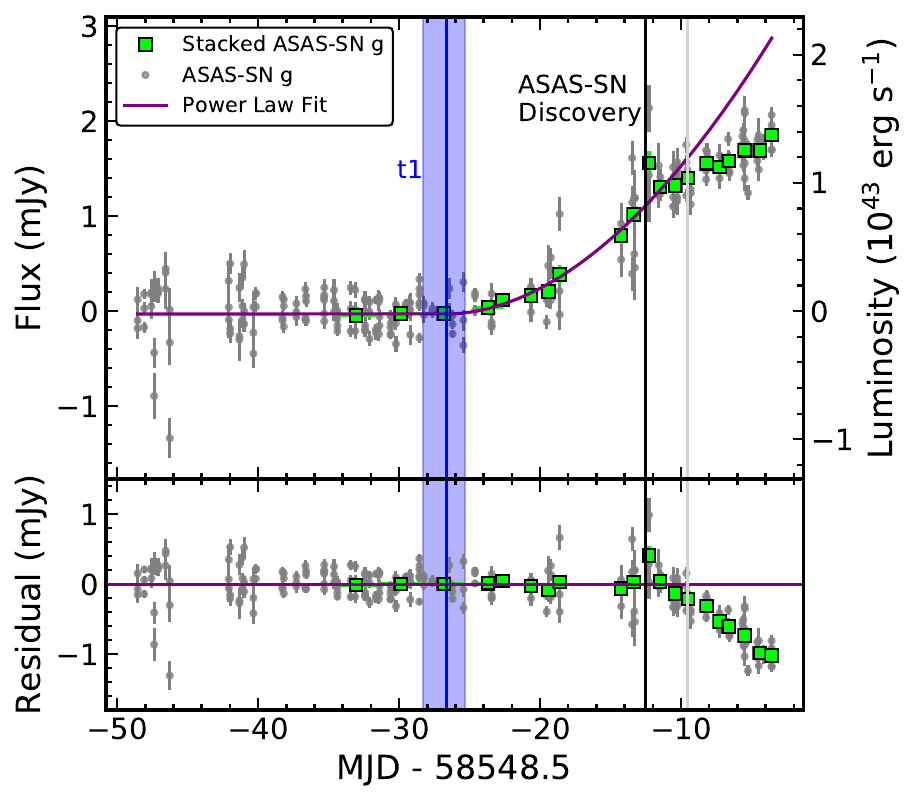}\hfill
 \caption{\textit{Top panel}: Stacked (green) and raw (gray) ASAS-SN $g$-band light curve and best-fit power-law model in purple. This power-law fit yields a time of first light of $t_1$ = $58521.9^{+1.3}_{-1.7}$ and a power-law index of $\alpha = 1.90^{+0.42}_{-0.36}$. The blue line shows the fitted time of first light with the shading representing the uncertainty and the black line shows the ASAS-SN discovery date. The light gray line marks the last epoch that was fit by our MCMC model. \textit{Bottom panel}: Residuals between the data and best-fit power-law model. The ASAS-SN $g$-band light curves deviates from a power-law rise roughly 16 days after the fitted time of first light.}
 \label{fig:rise}
\end{figure}

ASASSN-19dj is one of only a few TDEs for which the early-time coverage is adequate to fit a rise slope. We fit the early-time rise as a power-law with
\begin{equation}
f = z \text{ for $t < t_1$, and}
\end{equation}
\begin{equation}
f = z + h\bigg(\frac{t - t_1}{\text{days}}\bigg)^\alpha \text{for $t > t_1$}
\end{equation}
This model fits for the zero point $z$, the time of first-light $t_1$, a flux scale $h$, and the power-law index $\alpha$. An MCMC fit yields the best fit parameters $z = -30.7^{+8.6}_{-9.2} \  \mu\text{Jy}, \ h = 7.4^{+14.0}_{-5.5} \ \mu\text{Jy}, \ t_1 \text{(MJD)} = 58521.9^{+1.3}_{-1.7}, \ \text{and} \ \alpha = 1.90^{+0.42}_{-0.36}$. These fits are shown in Figure \ref{fig:rise}. 

ASASSN-19dj is only the third TDE for which a power-law could be fit to the early-time light curve. This best-fit power-law index of $\alpha = 1.90^{+0.42}_{-0.36}$ is consistent with the fireball model used for the early-time evolution of SNe \citep[e.g.][]{riess99, nugent11}. ASASSN-19bt, the TDE with the best early-time data also has a rise consistent with this model \citep{holoien19c}. Unlike the model invoked for SNe, where the ejecta initially expands at a constant velocity and temperature, the early stages of a TDE are more complex, so it is somewhat odd that these two objects have shown such a rise. Further analysis of more early-time TDE light curves will help us better understand their rise slopes. Additionally, the rise slope of the early-time bolometric light curve of ZTF19abzrhgq/AT2019qiz was found to be similar to ASASSN-19dj and ASASSN-19bt \citep{nicholl20}. It has been suggested that a $t^2$ rise may be the result of an outflow \citep[e.g.][]{nicholl20}. The existence of blue-shifted emission lines early in the evolution of ASASSN-19dj may support this possibility. Finally, the ASAS-SN $g$-band light curve follows a $t^2$ power-law rise for approximately 16 days, similar to ASASSN-19bt \citep{holoien19c}.

From Figure \ref{fig:rise}, we see that the light curve rises from the time of first light to the peak UV/optical bolometric luminosity in 26 days, shorter than the rise to peak time measured for ASASSN-19bt \citep{holoien19c} and the limits on rise times for PS18kh \citep{holoien19b}, and ASASSN-18pg \citep{holoien20}. Assuming that the early-time X-rays are due to accretion and are evidence of prompt circularisation (see Section \ref{sec:xray}), this more rapid rise may indicate a more efficient circularisation of material for ASASSN-19dj than other TDEs. From the fitted time of first light, we find that ASAS-SN discovered this transient within about two weeks of the beginning of the flare.

We used the Modular Open-Source Fitter for Transients \citep[\texttt{MOSFiT};][]{guillochon17b, mockler19} to fit the host-subtracted light curves of ASASSN-19dj to estimate physical parameters of the star, SMBH, and the encounter. \texttt{MOSFiT} uses models containing several physical parameters to generate bolometric light curves of a transient, generates single-filter light curves from the bolometric light curves, and fits these to the observed multi-band data. It then finds the combination of parameters yielding the highest likelihood match for a given model using one of various sampling methods. We ran the \texttt{MOSFiT} TDE model in nested sampling mode when fitting our data, as we have a large number of observations in several photometric filters.

\begin{figure*}
\begin{minipage}{\textwidth}
\centering
{\includegraphics[width=0.95\textwidth]{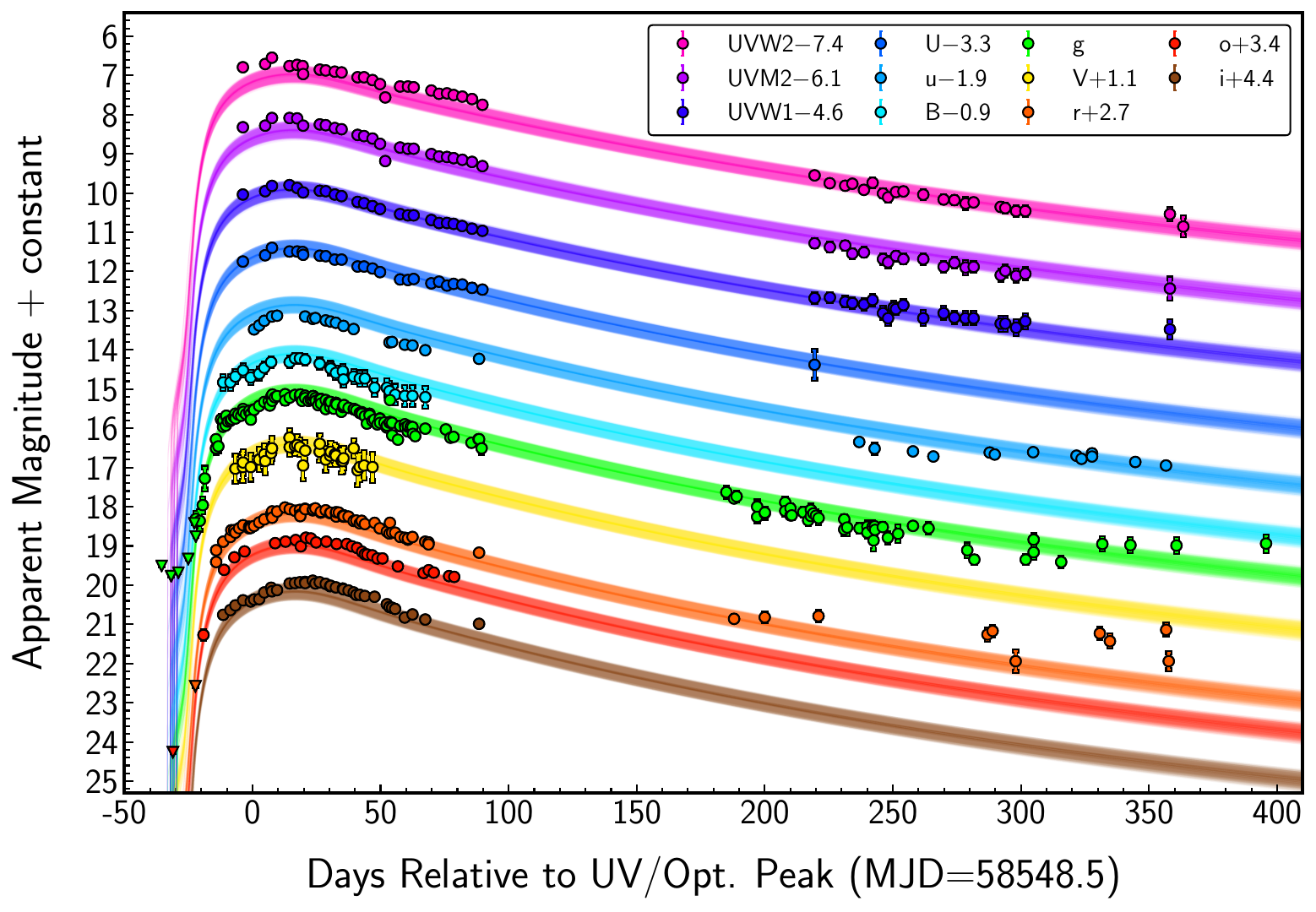}}
\caption{\texttt{MOSFiT} light curve fits and host-subtracted light curves. The the $1-99$\% range of fits for each filter are shown as shaded regions with the median fit shown as a solid line. All detections are plotted as circles with $3\sigma$ upper limits plotted as downward triangles. The colours match those of Figure~\ref{fig:opt_lc}.}
\label{fig:mosfit_lc}
\end{minipage}
\end{figure*}

\texttt{MOSFiT} is the only available tool for generalised fitting of TDE emission, and seems to model cases such as ASASSN-19dj, which has relatively smooth light curves, fairly well. We show the \texttt{MOSFiT} multi-band fits to the ASASSN-19dj light curves in Figure~\ref{fig:mosfit_lc} with our data overplotted. \texttt{MOSFiT} does a reasonable job of fitting the data near the peak, particularly in the optical, though it slightly underestimates the UV emission near peak. The rise is particularly well constrained compared to many of the TDEs in the sample fit by \citet{mockler19}.The fit to the late-time UV data is good, but many of the optical bands appear to flatten relative to the \texttt{MOSFiT} decline. The UV excess near peak and the late-time optical flattening may indicate the presence of multiple emission components throughout the evolution of ASASSN-19dj. 

\begin{table}
\centering
 \caption{\texttt{MOSFiT} TDE Model Parameter Fits}
 \label{tab:mosfit_params}
 \begin{tabular}{lrc}
  \hline
  Quantity & Value & Units\\
  \hline
  $\log{R_{\textrm{ph0}}}$ (photosphere power law constant) & $0.23^{+0.43}_{-0.42}$& --- \\
  $\log{T_{\textrm{viscous}}}$ (viscious delay timescale) & $-0.09^{+0.39}_{-0.56}$ & days \\
  $b$ (scaled impact parameter $\beta$) & $0.99^{+0.25}_{-0.93}$ & --- \\
  $\log{M_{h}}$ (SMBH mass) & $6.89^{+0.22}_{-0.23}$& \msun \\
  $\log{\epsilon}$ (efficiency) & $-0.44^{+0.71}_{-0.70}$ & --- \\
  $l$ (photosphere power law exponent) & $1.84^{+0.27}_{-0.27}$ & --- \\
  $\log{n_{\textrm{H,host}}}$ (local hydrogen column density) & $20.71^{+0.02}_{-0.02}$ & cm$^{-2}$ \\
  $M_\star$ (stellar mass) & $0.10^{+0.37}_{-0.08}$ & \msun \\
  $t_\textrm{exp}$ (time of disruption) & $-9.09^{+15.79}_{-15.86}$ & days \\
  $\log{\sigma}$ (model variance) & $-0.68^{+0.01}_{-0.01}$ & --- \\
  \hline
 \end{tabular}\\
\begin{flushleft}Best-fit values and $1-99$\% ranges for the \texttt{MOSFiT} TDE model parameters. Units are listed where appropriate. The listed uncertainties include both statistical uncertainties from the fit and the systematic uncertainties listed in Table 3 of \citet{mockler19}. \end{flushleft}
\end{table}

Table~\ref{tab:mosfit_params} shows the median values and $1-99$\% range for the \texttt{MOSFiT} TDE model parameters. The model parameters are generally very well constrained, with statistical uncertainties from the fit being significantly smaller than the systematic uncertainties of the model \citep[see Table 3 of][]{mockler19}. The black hole mass and stellar mass given by \texttt{MOSFiT} are $M_h=7.8^{+3.9}_{-4.1}\times10^6$~\msun and $M_\star=0.10^{+0.37}_{-0.08}$~\msun, respectively. This black hole mass is larger than, but marginally consistent with, the mass limit calculated by \citet{vanvelzen19} and consistent with our estimate in \S \ref{sec:archival}. The stellar mass, while low, is consistent with several other TDEs modelled in \citet{mockler19}. Finally, \texttt{MOSFiT} indicates that the star was likely completely disrupted in the encounter, though the lower limit on the scaled $\beta$ parameter $b$ is consistent with a partial disruption when the systematic uncertainties are taken into account.

To understand peak emission in the context of stream-stream collisions, we use \texttt{TDEmass} \citep{ryu20c}. \texttt{TDEmass} assumes that the UV/optical emission is shock-powered and extracts the SMBH and stellar mass based on the observed peak luminosity and temperature at peak. Using our peak luminosity of $(6.14 \pm 0.17) \times 10^{44}$ erg s$^{-1}$ and temperature at peak of $50400^{+3800}_{-6300}$ K, we obtain a SMBH mass of  $7.3^{+2.4}_{-1.0} \times 10^5$ \msun and a disrupted stellar mass of $13.0 _{-2.0}^{+4.2}$ \msun, both consistent with the values obtained by \citep{ryu20c}. Given the stellar mass of the host galaxy, this SMBH mass is roughly 15 times lower than expected from scaling relations. The stellar mass is unlikely for any TDE and inconsistent with the absence of recent star formation in the host galaxy.

\subsection{Spectra}  \label{sec:spectra}

\begin{figure*}
\centering
 \includegraphics[width=0.95\textwidth]{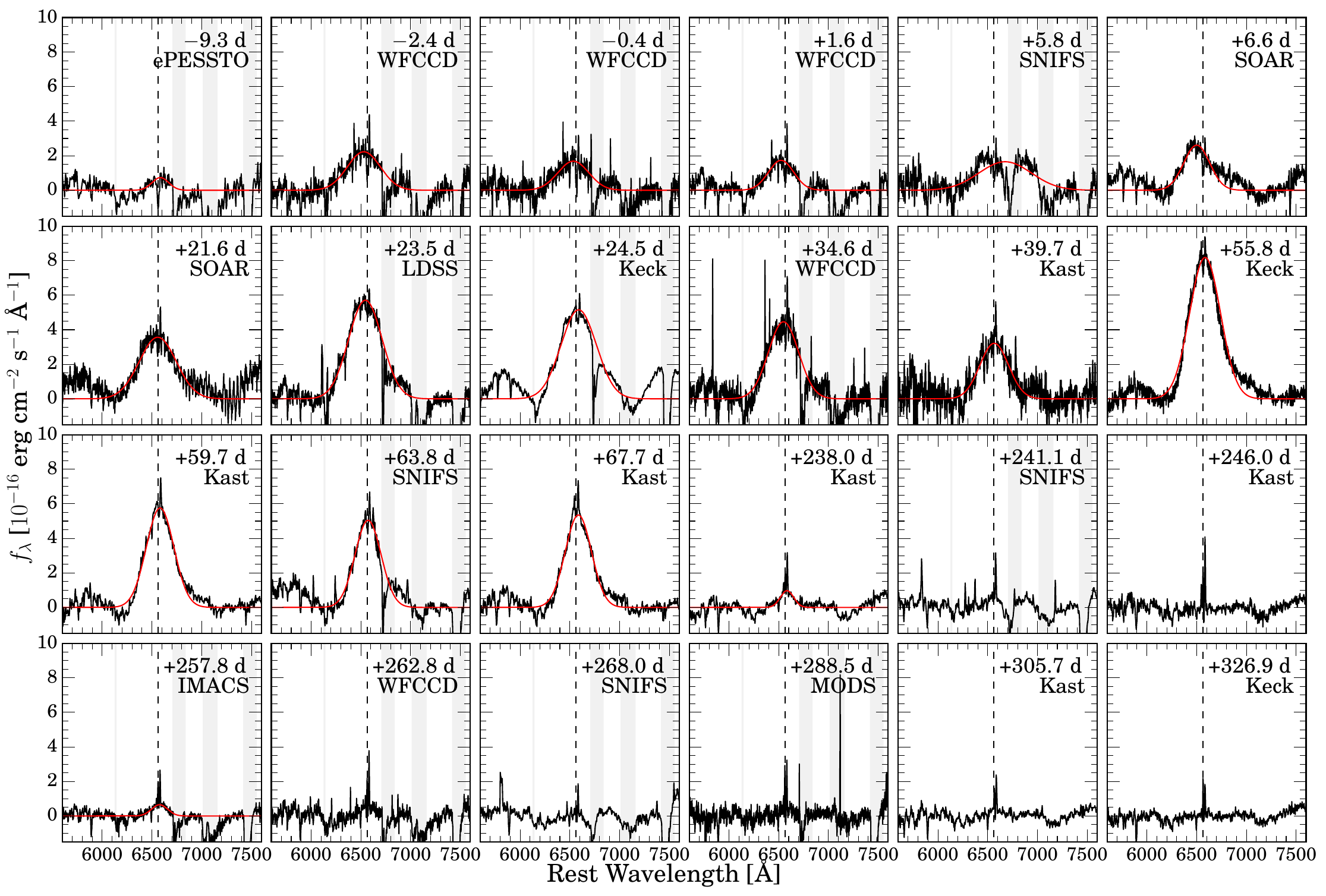}
 \caption{Spectral evolution of the H$\alpha$ line, increasing in time from top left to bottom right. The days relative to peak (MJD = 58548.5) in observer-frame days, of that particular spectral epoch and the instrument used to take the spectrum are shown in each individual panel. The red solid lines are Gaussian fits to the H$\alpha$ line profile, and are only shown for epochs with evidence of line emission. The vertical gray bands mark masked atmospheric telluric features.}
 \label{fig:halpha_evol}
\end{figure*}

\begin{figure}
\centering
 \includegraphics[width=.46\textwidth]{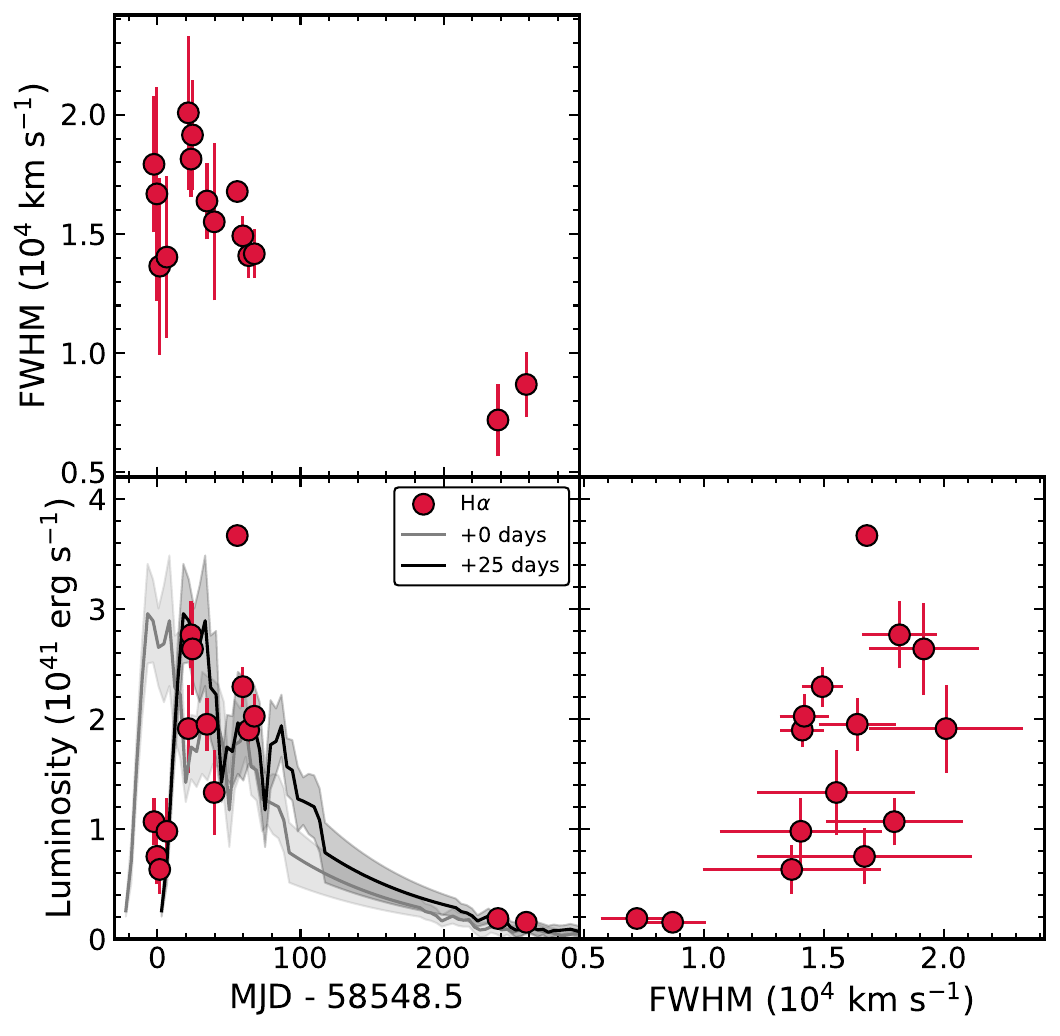}\hfill
 \caption{\textit{Top Left Panel}: H$\alpha$ FWHM as a function of time. \textit{Bottom Left Panel}: Time evolution of H$\alpha$ luminosity (red points). The bolometric light curve is scaled by a factor of $4.5 \times 10^{-4}$ and shown with shaded error bars at zero days offset (gray) and 25 days offset (black) to highlight the delay in bolometric and H$\alpha$ peak. The bolometric light curves are smoothed by linearly interpolating to a time-series with the same length as the original coverage, but with half the number of points. \textit{Right Panel}: H$\alpha$ luminosity as compared to the FWHM of the line.}
 \label{fig:fwhm}
\end{figure}

The early-time spectra of ASASSN-19dj have the very blue continuum that is a hallmark of tidal disruption events. This excess in blue flux grows towards peak light and fades back to host galaxy levels at later times. The very early-time optical spectra of ASASSN-19dj lack many of the spectral features that TDEs usually exhibit. For example, in the earliest spectrum, taken approximately 17 days after first light (see \S \ref{lc}), there are no clear strong broad H and He lines. 

We traced the evolution of the prominent H$\alpha$ feature by subtracting a linear continuum normalized in the regions around 6200~\AA\ and 7200~\AA. We then modeled the continuum-subtracted H$\alpha$ profile as a Gaussian. We masked narrow H$\alpha$+[\ion{N}{ii}] lines for all spectra and telluric absorption features for spectra without telluric corrections. Looking at Figure \ref{fig:halpha_evol}, which zooms into a small region around H$\alpha$, we see that the H$\alpha$ line slowly grows in strength from MJD $\simeq$ 58539 to MJD $\simeq$ 58544, and quickly becomes very strong by MJD $\simeq$ 58571, with a peak line flux of $\sim 3.7 \times 10^{-13} \text{erg cm}^{-2} \text{ s}^{-1}$ or a luminosity of $\sim 4.3 \times 10^{41} \text{erg} \text{ s}^{-1}$ at the distance of ASASSN-19dj. The line luminosity remains roughly constant until at least 41 days later, after which the source became Sun-constrained. After the seasonal gap, we find evidence for weak H$\alpha$ emission as late as $\sim 260$ days after peak, consistent with other optical TDEs \citep[e.g.][]{hung20, holoien20}. 

Figure \ref{fig:fwhm} shows the luminosity and FWHM evolution of H$\alpha$ as a function of time and the luminosity versus line width. The line width is relatively constant prior to peak line flux and decreases thereafter, with some epochs exhibiting very broad H$\alpha$, up to roughly $2 \times 10^4 \text{km s}^{-1}$. There appears to be a positive correlation between H$\alpha$ line flux and line width. To test this, we performed the Kendall Tau test and find $\tau = 0.49$ and a corresponding p-value of 0.01 indicating a significant moderately strong correlation. The positive correlation seen here between H$\alpha$ line flux and line width agrees with what has been observed in the TDEs PS18kh \citep{hung19, holoien19c}, ASASSN-18pg \citep[][]{leloudas19, holoien20}, and ASASSN-18ul \citep[][]{wevers19}. Unlike other TDEs such as ASASSN-18pg \citep[][]{holoien20}, we do not see evidence for both a broad and narrow component of the H$\alpha$ emission. 

\begin{figure*}
\centering
 \includegraphics[width=0.95\textwidth]{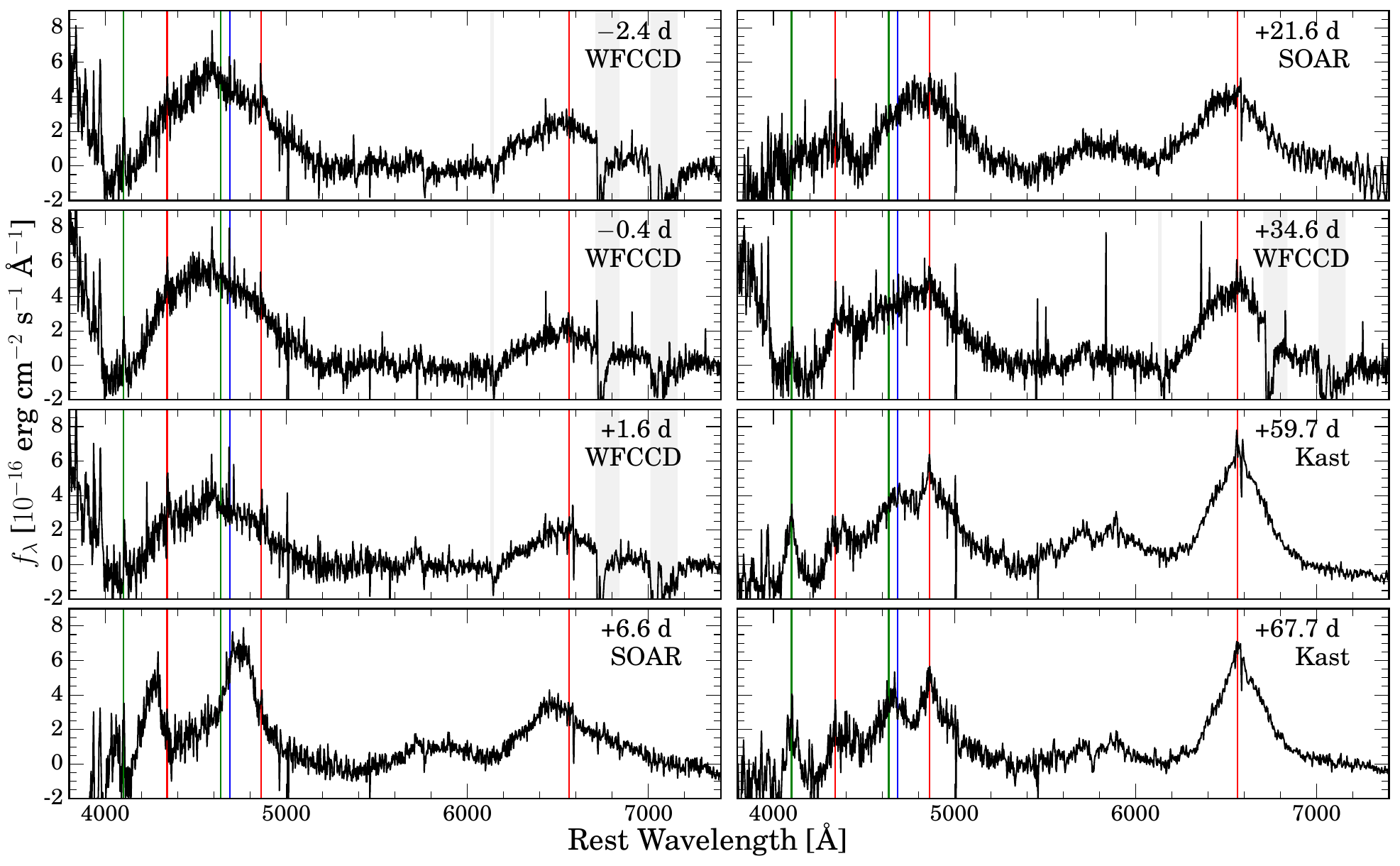}
 \caption{Host-subtracted spectral evolution of ASASSN-19dj increasing in time from top left to bottom right. The days relative to peak (MJD = 58548.5) in observer-frame days, of that particular spectral epoch and the instrument used to take the spectrum are shown in each individual panel. Vertical lines mark emission features common in TDEs with red marking H, blue marking \ion{He}{ii} and green marking \ion{N}{iii} lines. The vertical gray bands mark masked atmospheric telluric features.}
 \label{fig:bluespec}
\end{figure*}

While the dominant spectral features of ASASSN-19dj appear to be broad hydrogen lines, \citet{vanvelzen20} classify ASASSN-19dj as a TDE-Bowen object. There may be some evidence for broad emission centered on $\sim 4600$ \AA, although the origin of this feature is difficult to determine. In order to analyze this feature in more detail, it was necessary to isolate the TDE emission flux. We first subtracted the host emission using the archival SDSS spectrum, then fit a Legendre polynomial to the continuum of the host-subtracted spectrum, and then subtracted the fitted continuum to get a host- and continuum-subtracted spectrum. The resulting spectra are shown in Figure \ref{fig:bluespec} with the spectra selected to highlight the early temporal evolution.

By the +6.6 day SOAR spectrum, there is clear evidence for broad, blueshifted H$\alpha$ and H$\beta$ emission. The line profile of H$\alpha$ in particular is similar to PS17dhz/AT2017eqx \citep{nicholl19} and ZTF19abzrhgq/AT2019qiz \citep{nicholl20}, suggestive of an outflow \citep{roth18}. Interestingly, the H$\alpha$ and H$\beta$ lines in the +6.6 and +21.6 day SOAR spectra appear to have flat tops, like PS18kh \citep{holoien19b} and ASASSN-18zj \citep{short20, hung20}, although for ASASSN-19dj this feature does not persist for long. Additionally, there is significant evolution in the broad features near 4600 \AA. In pre-peak spectra, there is a very broad feature, similar to the early broad \ion{He}{ii} lines of ASASSN-15oi \citep{holoien16c}, which quickly change to distinct H lines. Similar to ZTF19abzrhgq/AT2019qiz, as the H lines return to their rest wavelengths the strengths of the Bowen fluorescence lines grow dramatically, with strong \ion{N}{iii} and \ion{He}{ii} emission by roughly 60 days after peak in addition to the now-dominant hydrogen emission.

Throughout the evolution of ASASSN-19dj, the H$\alpha$ line is very broad. There may be weak evidence for a broadening of the H$\alpha$ line at early times, similar to PS18kh \citep[][]{holoien19b}, although these epochs have large uncertainties on the FWHM. The H$\alpha$ line begins to narrow again after the peak H$\alpha$ luminosity. There is a time delay between the peak UV/optical magnitude, which occurs at MJD = 58545.5, and the peak H$\alpha$ luminosity, at roughly MJD $\simeq$ 58571. As can be seen in Figure \ref{fig:fwhm}, the H$\alpha$ luminosity tracks the bolometric luminosity reasonably well when offset by 25 days. While this delay of $\sim 25$ days is only approximate, given our lack of spectra near this time, this provides an upper-limit on how far from the SMBH this emission is located. If the lines are due to reprocessing of high-energy FUV and X-ray photons produced in an accretion disc, this suggests the existence of reprocessing material at a distance of several tens of thousands of gravitational radii, slightly larger than the distance to reprocessing material derived for ASASSN-18pg, which showed clear evidence for a delay between the bolometric peak and peak line emission \citep{holoien20}. The early-time H$\alpha$ line profiles, with a blueshifted core and asymmetric red wing support the idea of an optically thick outflow as the reprocessing material \citep{roth18} as do the existence of early-time He emission without corresponding H emission \citep{roth16, roth18}.

\subsection{Spectral Energy Distribution} \label{sed}

\begin{figure}
\centering
 \includegraphics[width=0.48\textwidth]{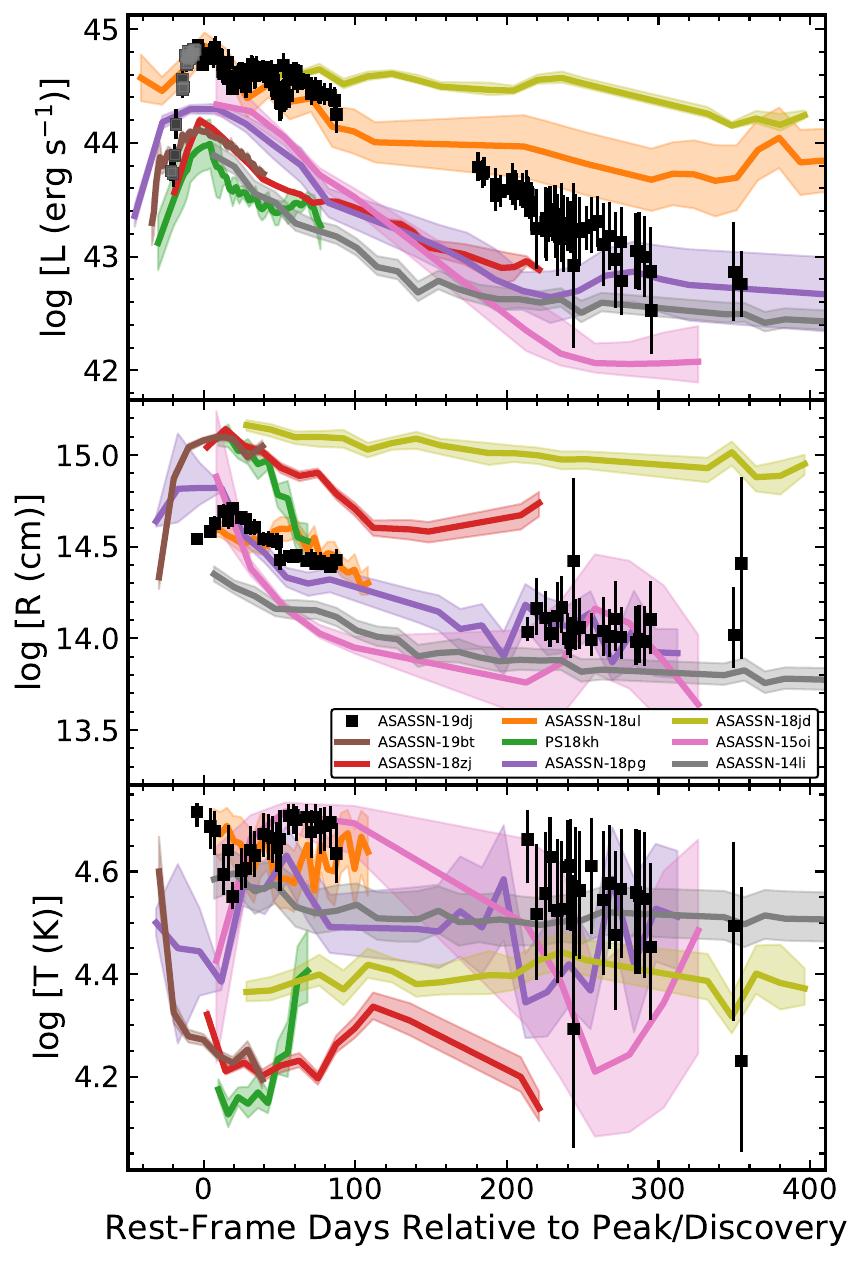}\hfill
 \caption{Evolution of the UV/optical blackbody luminosity (top panel), radius (middle panel), and temperature (bottom panel) for ASASSN-19dj (black squares), in comparison to other TDEs with strong H$\alpha$ emission, strong X-ray emission, or both: ASASSN-19bt \citep[brown line;][]{holoien19c}, ASASSN-18zj \citep[red line;][]{vanvelzen20}, ASASSN-18ul \citep[orange line;][Payne et al., in preparation]{wevers19}, PS18kh \citep[green line;][]{holoien19b}, ASASSN-18pg \citep[purple line;][]{holoien20}, ASASSN-15oi \citep[pink line;][]{holoien18a}, and ASASSN-14li \citep[gray line;][]{brown17a} in addition to the TDE/AGN ASASSN-18jd \citep[gold line;][]{neustadt19}. The lines are smoothed over the individual epochs by linearly interpolating to a time-series with the same length as the original coverage, but with half the number of points. Time is in rest-frame days relative to the peak luminosity for the objects discovered prior to peak (ASASSN-19dj, ASASSN-19bt, ASASSN-18zj, ASASSN-18ul, PS18kh, and ASASSN-18pg), and reative to discovery for those which were not (ASASSN-18jd, ASASSN-15oi, and ASASSN-14li). The gray squares for ASASSN-19dj indicate where data has been bolometrically corrected using the ASAS-SN $g$-band light curve assuming the temperature from the first \swift epoch was constant.}
 \label{fig:BB_fit}
\end{figure}

Figure \ref{fig:BB_fit} shows the blackbody model fits in terms of luminosity, radius, and temperature for ASASSN-19dj. ASASSN-19dj is one of the most luminous TDEs discovered to date, with a peak luminosity of $(6.15 \pm 0.17) \times 10^{44} \text{ erg s}^{-1}$, consistent with the value derived by \citet{liu19}. This peak luminosity is comparable only to the TDE ASASSN-18ul \citep[][Payne et al. in preparation]{wevers19} and the TDE/AGN ASASSN-18jd \citep{neustadt19}. The decline in bolometric luminosity of ASASSN-19dj is quite slow, which is consistent with the findings of \citet{hinkle20a} that more luminous TDEs decay slower than less luminous TDEs. At later times (over $\sim 250$ days after peak) the luminosity appears to flatten out, consistent with other TDEs with late-time observations including ASASSN-14li \citep{brown17a}, ASASSN-15oi \citep{holoien16c}, ASASSN-18pg \citep{leloudas19,holoien20}, ASASSN-18ul \citep[][Payne et al. in preparation]{wevers19}, and ATLAS18way \citep{vanvelzen20}. In the cases of both ASASSN-19dj and ASASSN-15oi \citep{gezari17, holoien18a}, the flattening of the bolometric light curve is roughly coincident with and increase in X-ray flux.

The blackbody radius of ASASSN-19dj is initially relatively small compared to other well-studied TDEs with similar strong H$\alpha$ emission such as ASSASN-18zj \citep{hung20, short20}, ASASSN-19bt \citep{holoien19c}, and PS18kh \citep{holoien19b} as well as the TDE/AGN ASASSN-18jd \citep{neustadt19}. It is larger however than other X-ray bright TDEs like ASASSN-14li \citep{holoien16a, brown17a} and ASASSN-15oi \citep{holoien16c}. At late times the radius continues to decrease slowly, becoming consistent with other TDEs with well-sampled late-time evolution such as ASASSN-14li \citep{holoien16a} and ASASSN-15oi \citep{holoien18a}. Additionally this slow late-time decrease in radius is consistent with many TDEs in the literature \citep[e.g.][]{vanvelzen20, hinkle20a}.

Unlike other TDEs with strong H$\alpha$ emission (ASASSN-18zj, ASASSN-19bt, and PS18kh) and the TDE/AGN ASASSN-18jd, the temperature of ASASSN-19dj is quite hot, on the order of $\sim 45000$ K. This temperature is more in line with the TDE-Bowen spectral class introduced by \citet{vanvelzen20}. This hot temperature, especially at late times, is similar to the TDEs ASASSN-18pg, ASASSN-15oi, ASASSN-14li, with the the latter two also exhibiting late time X-ray emission. The blackbody temperatures of each of these TDEs are mostly flat throughout the evolution of the TDE. The blackbody temperature of ASASSN-19dj appears to decrease near peak, which can also be seen in Figure \ref{fig:opt_lc} as the time of peak is earliest in the bluest bands and delayed in each of the red bands, similar to other TDEs with high-cadence pre-peak photometry such as ASASSN-19bt \citep{holoien19c} and ASASSN-18pg \citep{holoien20}.

\begin{figure}
\centering
 \includegraphics[width=.48\textwidth]{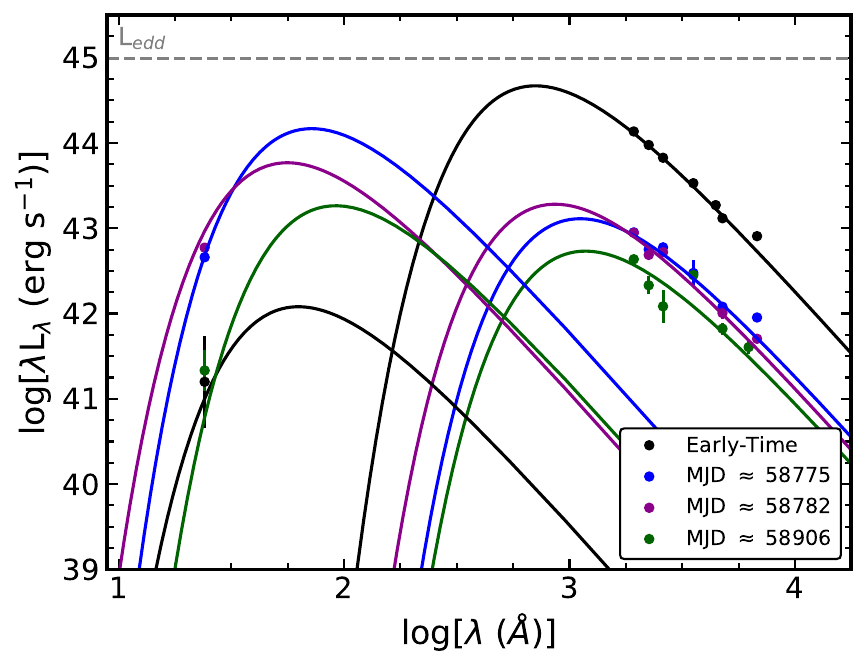}\hfill
 \caption{Spectral energy distribution of ASASSN-19dj at four different epochs. The first epoch (black) is at MJD = 58544 for the UV/optical emission and is the stacked early-time \swift for the X-ray. The second epoch (blue) is the first epoch after the X-ray brightening at MJD $\approx$ 58775 for both the UV/optical and X-ray. The third epoch (purple) is the epoch of peak X-ray emission at MJD $\approx$ 58782. The fourth epoch (green) is a late-time epoch at MJD $\approx$ 58906. For each epoch, the data is shown as points while the lines represent the best fit blackbody components for the UV/optical and X-ray emission. The dashed gray line indicates the Eddington luminosity for an SMBH of mass $7.8 \times 10^6 \msun$, the SMBH mass derived from \texttt{MOSFiT}.}
 \label{fig:sed}
\end{figure}

The UV/optical and X-ray SEDs at four epochs in the evolution of ASASSN-19dj are shown in Figure \ref{fig:sed}. The UV/optical emission of ASASSN-19dj is dominant at early times and several orders of magnitude brighter than the X-ray emission. Using the SMBH mass derived from \texttt{MOSFiT}, we calculate an Eddington luminosity of $9.8 \times 10^{44}$ erg s$^{-1}$. Similar to what we find from the X-ray properties, we find no clear evidence for a plateau caused by Eddington-limited accretion near peak in either the bolometric light curve (see Fig. \ref{fig:BB_fit}) or the single-band UV and optical light curves (see Fig. \ref{fig:opt_lc}), although we caution that there is significant scatter in the SMBH mass estimates for this source. The Eddington ratio for the peak bolometric luminosity is $\sim 0.6$, which is consistent with other UV/optical TDEs in the literature \citep[e.g.][]{wevers17, mockler19}. By roughly 250 days after peak, the X-ray emission exceeded the UV/optical emission by roughly an order of magnitude, similar to the late-time X-ray brightening of ASASSN-15oi \citep{gezari17, holoien18a}, as well as the flatter late-time X-ray emission of ASASSN-14li \citep{brown17a}. During the epoch of peak X-ray emission, at MJD $\approx 58782$, the difference between the UV/optical and X-ray SEDs is less pronounced. By roughly a year after peak, both the X-ray and UV/optical SEDs have faded and are comparable in peak luminosity.

\subsection{Pre-ASASSN-19dj Outburst?}
The CRTS light curve of the host galaxy, KUG 0810+227, shows evidence of a previous outburst in September 2005 (see Figure \ref{fig:crts}), roughly 14.5 years prior to this TDE. The data quality of the archival CRTS images of the host galaxy were too poor to perform image subtraction. Instead, we stacked six CRTS images of the host galaxy during the outburst and seventeen references images taken at least five years after the outburst. Through comparison of the image centroids for these stacks, we find the difference to be 0\farcs{18}$\pm$0\farcs{32}, corresponding to a physical distance of $84 \pm 152$ pc. Given the low quality of the archival images, this uncertainty is estimated by taking the standard deviation of the centroids of each of the individual images. While this constraint on the location of this previous transient is based on unsubtracted images, and therefore includes host light, it appears to be consistent with the host nucleus. However, we can make several statistical statements based on previous analysis presented in this paper.

The absolute magnitude of the brightest CRTS epoch is $V = -19.1$ or $L_V = 1.4 \times 10^{43}$ erg s$^{-1}$ (corrected for Galactic extinction, but assuming no host galaxy reddening), which is more luminous than the observed magnitudes of many types of supernovae, but consistent with the absolute magnitudes of Type Ia supernovae \citep[e.g.]{folatelli10, richardson14}. We attempted to fit the CRTS light curve of the archival outburst with SNooPy \citep{burns11} to constrain the properties of the light curve. We used the default E(B$-$V) model, but assumed no host galaxy reddening given the single filter light curve. These fits get their shape information from the fitted $\Delta m_{15}$ and use the K-corrections of \citet{hsiao07}, the Milky Way dust map of \citet{schlegel98}, and SN templates of \citet{prieto06}. From this, we find that the decline of this outburst is somewhat slower than expected for a Type Ia supernova, with $\Delta m_{15} = 0.8 \pm 0.4$ mag. Yet, given the data quality and maximum observed $V$-band magnitude, we cannot rule out a luminous Type Ia SN. The true peak luminosity of this transient is likely higher than $1.4 \times 10^{43}$ erg s$^{-1}$, because a seasonal gap occurred immediately prior to the CRTS detection. Even if the peak luminosity is higher than this it may still be consistent with the tail of observed SN Ia magnitudes \citep[e.g.][]{folatelli10, richardson14} or a superluminous supernova.

Next, we evaluated the possibility that this CRTS flare was a previous TDE, by estimating a TDE rate for the host galaxy. The rate of TDEs is roughly $10^{-4}$ -- $10^{-5} \text{ yr}^{-1}$ per galaxy \citep[e.g.][]{vanvelzen14, holoien16a, auchettl18} for an average galaxy. However, KUG 0810+227 is a post-starburst galaxy, for which it is known that the TDE rate can be enhanced by up to 200 times the average \citep[e.g.][]{french16, law-smith17, graur18}, and thus it would not be unreasonable that a TDE could occur every 50 - 500 years. Even within this sample of post-starburst galaxies, KUG 0810+227 appears to be extreme in terms of its Lick H$\delta_A$ index. Counting galaxies with H$\delta_{A} - \sigma$(H$\delta_{A}) >$ 7.0 \AA\ and H$\alpha$ emission EW $<$ 3.0 \AA, gives just 0.025\% of all the galaxies in SDSS and a similar TDE rate enhancement of $\sim250$ times the average. Thus, there is the possibility that the CRTS flare is a previous TDE.

Several pieces of archival data and optical emission line diagnostics are consistent with KUG 0810+227 being a LLAGN. The line ratios of the optical spectrum of KUG 0810+227 lie in the Seyfert region of two line ratio diagnostic diagrams, which suggests the possibility of the host being a LLAGN. However, we note that the WHAN diagram classifies KUG 0810+227 as an RG, suggesting a possible non-AGN ionization source. In the X-ray, the first \swift XRT epoch gives a deep upper limit, which is consistent with a small fraction of observed X-ray luminosities of AGN \citep[e.g.][]{tozzi06, ricci17}. Therefore, because the host is consistent with a LLAGN, the previous flare could be associated with a pre-flare AGN outburst.

\subsection{X-rays} \label{sec:xray}

\begin{figure}
\centering
 \includegraphics[width=0.48\textwidth]{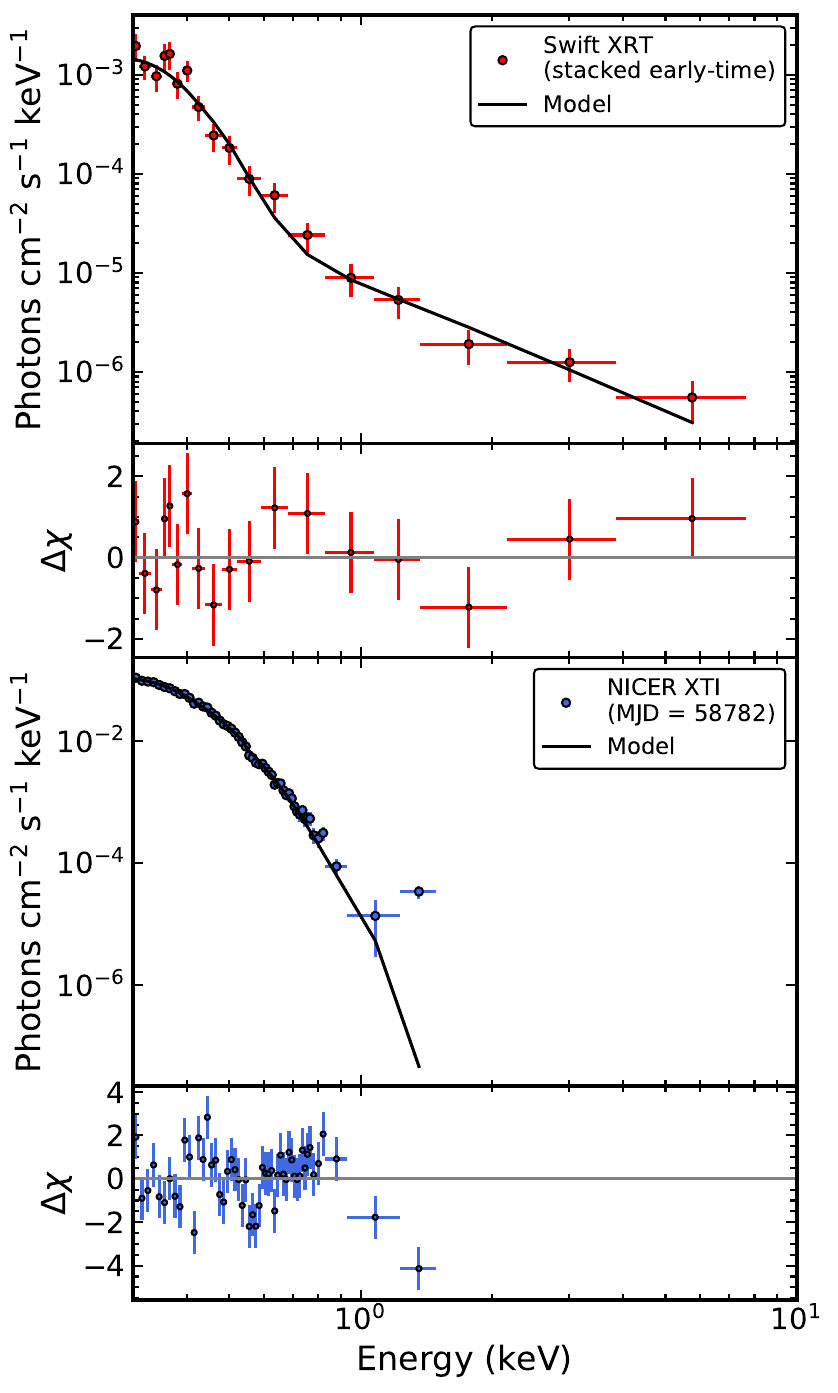}
 \caption{X-ray spectrum and $\Delta \chi$ for the stacked early-time \swift observations (top panel) and \textit{NICER} observations of ASASSN-19dj at peak X-ray emission on MJD = 58782 (bottom panel). The black lines in the top and bottom panels are the best-fit absorbed blackbody model + power-law and absorbed blackbody models respectively. The reduced $\chi^2$ of the best-fit models are 0.90 for 15 degrees of freedom and 1.72 for 50 degrees of freedom respectively.}
 \label{fig:xray_spectrum}
\end{figure}

\begin{table*}
\centering
 \caption{X-ray spectral parameters of ASASSN-19dj}
 \label{tab:xray_spec}
 \begin{tabular}{cccccccc}
  \hline
  MJD & $N_{H}$ & $N_{H}$ Error & kT & kT Error & log Radius & Radius Error & Satellite\\
  & ($10^{20}$ cm$^{-2}$) & ($10^{20}$ cm$^{-2}$) & (keV) & (keV) & (cm) & (cm) &\\
  \hline
  58544.75-58638.27 & 4.06 & 4.00 & 0.050 & 0.009 & 11.15 & 0.38 & \swift \\
  58767.93-58779.88 & 7.55 & 5.00 & 0.048 & 0.010 & 12.12 & 0.42 & \swift \\
  58782.67 & 3.98 & 3.00 & 0.062 & 0.005 & 11.69 & 0.29 & \swift \\
  \ldots & \ldots & \ldots & \ldots & \ldots & \ldots & \ldots & \ldots \\
  58920.19 & 4.16 & --- & 0.035 & 0.002 & 12.22 & 0.27 & \textit{NICER} \\
  58934.02 & 4.16 & --- & 0.030 & 0.002 & 12.54 & 0.33 & \textit{NICER} \\
  58940.53 & 4.16 & --- & 0.028 & 0.002 & 12.59 & 0.34 & \textit{NICER} \\
  \hline
 \end{tabular}\\
\begin{flushleft}Neutral hydrogen column densities, blackbody temperatures, and effect blackbody radii derived from the various X-ray spectral epoch. A range of MJD in the first column indicates the beginning and end of the range over which data were stacked to increase S/N. The last column reports the source of the data for each epoch. Only a small section of the table is displayed here. The full table can be found online as an ancillary file.\end{flushleft}
\end{table*}

\begin{figure*}
\centering
 \includegraphics[width=1.0\textwidth]{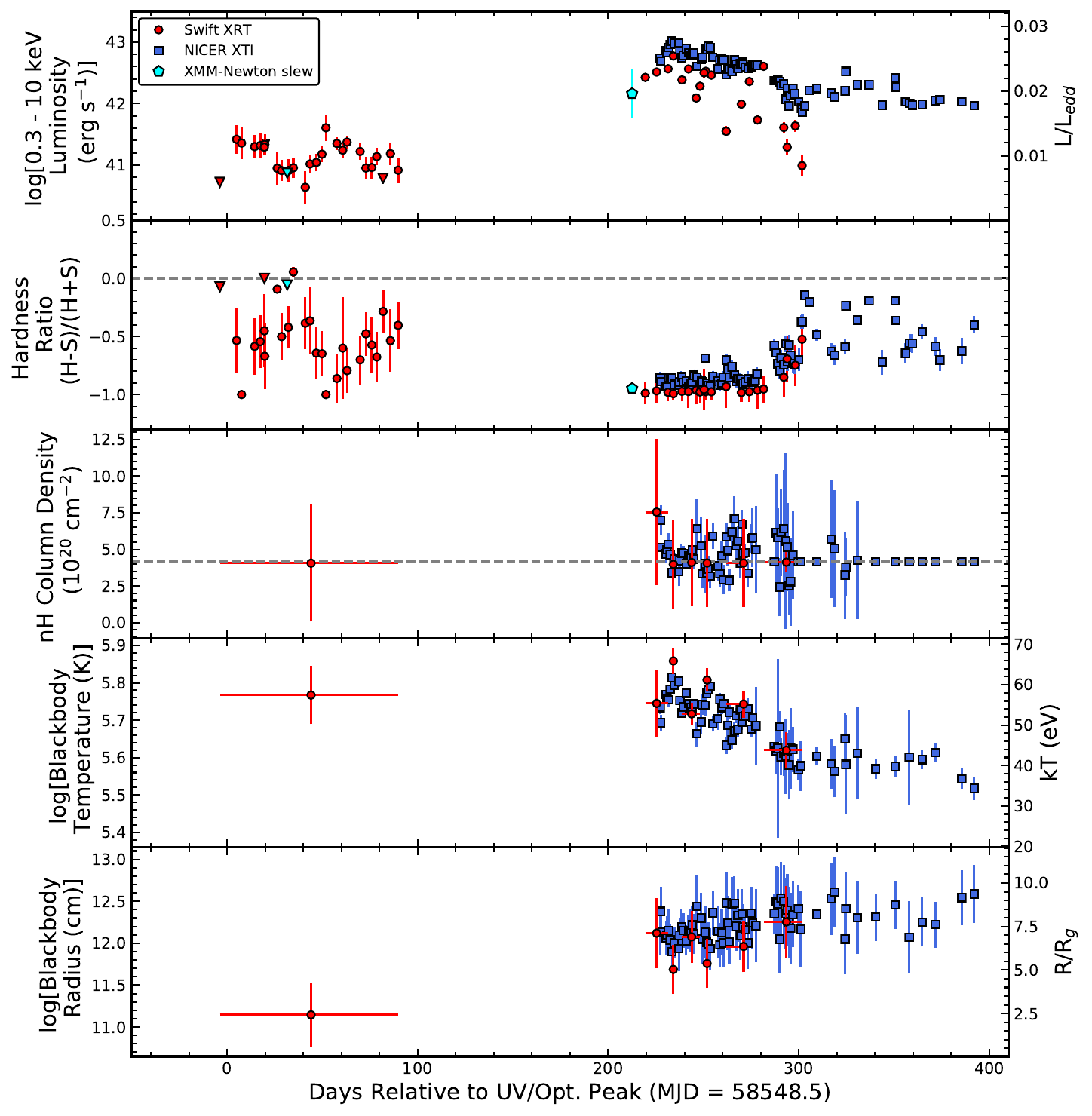}
 \caption{In order from top to bottom: X-ray luminosity, hardness ratio, neutral hydrogen column density (with the dashed gray line marking the Galactic column density), blackbody temperature, and blackbody radius of ASASSN-19dj measured with \textit{Swift} (red circles), \textit{NICER} (blue squares) and \textit{XMM-Newton} slew (cyan pentagon). We define hard counts H as the number of counts in the 2-10 keV range and soft counts S are the number of counts in the 0.3-2 keV range, with a gray dashed line marking zero. The hardness ratio is defined as (H$-$S)/(H+S). Downward-facing triangles mark upper limits.}
 \label{fig:all_xray}
\end{figure*}

ASASSN-19dj is one of several optical TDEs to show strong X-ray emission. In Figure \ref{fig:xray_spectrum}, we show the stacked early-time \swift spectrum and a \textit{NICER} spectrum from the epoch of peak X-ray emission. In Figure \ref{fig:all_xray} (top panel), we show the X-ray light curve as derived from both the individual \textit{Swift} and \textit{NICER} observations. To estimate the X-ray luminosity, we converted the extracted count rate into flux using WebPIMMS\footnote{\url{https://heasarc.gsfc.nasa.gov/cgi-bin/Tools/w3pimms/w3pimms.pl}} and assumed an absorbed blackbody model with a temperature of $\sim$50 eV, corresponding to the average blackbody temperature derived from our \textit{Swift} and \textit{NICER} X-ray spectra. This value is also consistent with the blackbody temperatures of other X-ray bright TDEs \citep[e.g. ASASSN-14li, ASASSN-15oi,][]{brown17a, holoien18a, kara18}. The first X-ray observation of ASASSN-19dj was taken using the \textit{Swift} XRT approximately 4 days before the peak UV/optical emission (MJD = 58544.8). During this observation, ASASSN-19dj showed no evidence of X-ray emission with a 3$\sigma$ upper limit of 6$\times10^{40}$ erg s$^{-1}$, consistent with the limits/detection of X-ray emission seen prior to peak in the TDEs ASASSN-19bt \citep{holoien19c}, ASASSN-18pg \citep[][]{leloudas19, holoien20}, ZTF19abzrhgq/AT2019qiz \citep{2019ATel13143....1A} and other X-ray TDE candidates \citep{auchettl17}. This upper limit places even stricter constraints on the possibility that the host galaxy is a LLAGN, with \citet{tozzi06} finding that fewer than $\sim10$\% of AGN have X-ray luminosities this low, and \citet{ricci17} measuring only 1\% of their unobscured non-blazar AGN sample to have X-ray luminosities this low. 

ASASSN-19dj was first detected in X-rays $\sim$9 days later, $\sim$4 days after the UV/optical peak, in the second \swift XRT observation, with its X-ray luminosity increasing by at least half an order of magnitude to $\sim3\times10^{41}$ erg s$^{-1}$. Similar to ASASSN-18jd \citep{neustadt19}, ASASSN-18ul \citep{wevers19} and ASASSN-15oi \citep{gezari17}, the X-ray emission of ASASSN-19dj showed significant variations in luminosity over the first $\sim$100 days after peak, varying between $\sim10^{40.7} - 10^{41.7}$ erg s$^{-1}$ before the seasonal gap, much larger than the variability seen in ASASSN-14li \citep{brown17a}, but similar to that seen in ASASSN-18jd \citep{neustadt19} or ASASSN-18ul \citep{wevers19}. Once the source became visible again $\sim$220 days after peak, \textit{XMM-Newton} slew, \textit{Swift} XRT, and \textit{NICER} observations found that the source had brightened by nearly a factor of $\sim$10. This brightening behaviour is reminiscent of what was seen in ASASSN-15oi \citep{gezari17, holoien18a} and hinted at in ASASSN-18ul \citep{wevers19}, in which the X-ray emission increased by an order of magnitude $\sim250$ days after peak brightness before fading. ASASSN-19dj peaked at an X-ray luminosity of $\sim10^{43}$ erg s$^{-1}$ before fading by nearly an order of magnitude over $\sim$100 days and then plateauing at an X-ray luminosity of $\sim10^{42}$ erg s$^{-1}$. The peak luminosity corresponds to an Eddington ratio between 0.01-0.03, consistent with other X-ray bright TDEs \citep{mockler19, 2019MNRAS.487.4136W} and again disfavoring Eddington-limited accretion as suggested by \citet{vanvelzen19}.

In Figure \ref{fig:all_xray} (second panel), we present the evolution of the X-ray hardness ratio\footnote{The hardness ratio (HR) is defined as HR = (H$-$S)/(H+S) where H is the number of counts in the 2.0-10.0 keV energy range and S is the number of counts in the 0.3-2.0 keV energy range} (HR) as a function of time. At early times, ASASSN-19dj shows significant variability in its hardness, varying between a soft HR of $-1$ and harder HR of $-0.2$ during the first 100 days. By 200-280 after days after peak, the hardness ratio of ASASSN-19dj plateaued to a soft HR between $-1$ and $-0.8$, before hardening significantly over $\sim20$ days from 280-300 days after peak. Finally, from 300 days after peak onwards, ASASSN-19dj returned to the HR variability observed at early times. The behaviour seen at early times is consistent with the presence of hard X-ray emission in the form of a power-law in addition to a soft thermal blackbody, consistent with \citet{liu19}. This can be seen in the merged \swift spectra derived from the observations taken within the first 100 days (see Fig. \ref{fig:xray_spectrum}). The softening of the X-ray emission between 200 to $\sim$280 days after discovery occurs when the X-ray emission from this event becomes dominated by a strong thermal blackbody component. Near peak, the constant HR with time and decreasing X-ray luminosity, is consistent with that exhibited by non-thermal TDEs such as ASASSN-14li \citep{auchettl18}. However, we note that the lack of significant HR evolution seen in ASASSN-14li begins at peak brightness in X-ray, UV/optical and bolometric luminosity and continues for thousands of days after peak. As the blackbody component cools with time and fades, the X-ray emission is seen to harden, similar to what was seen in ASASSN-14li \citep[e.g.][]{2018MNRAS.474.3593K}.  

\begin{figure}
\centering
 \includegraphics[width=0.48\textwidth]{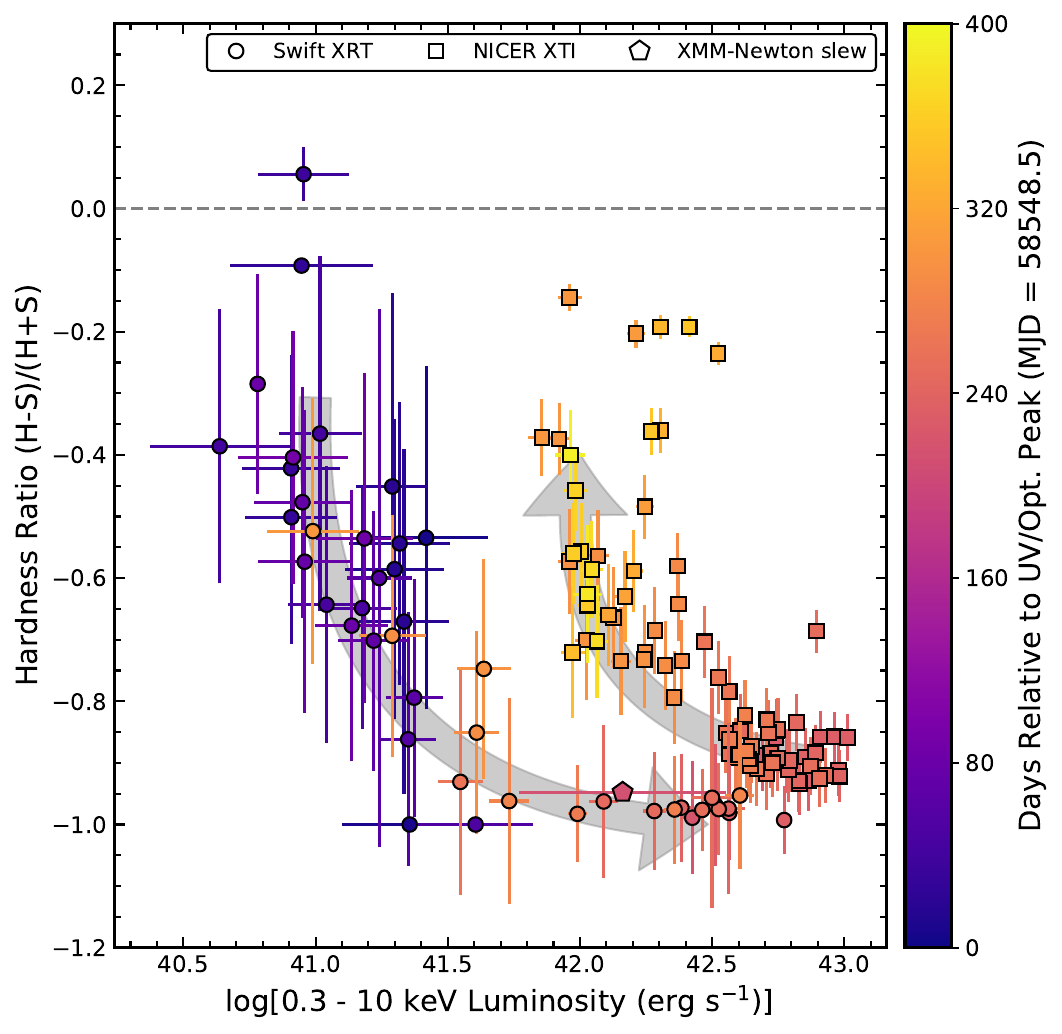}
 \caption{Hardness ratio as a function of X-ray luminosity as measured by \textit{Swift} (circles), \textit{NICER} (squares), and \textit{XMM-Newton} slew (pentagon) for the detections only. Light gray arrows indicate that as ASASSN-19dj becomes brighter, the X-ray emission becomes softer. This behaviour is similar to that seen in X-ray bright AGN \citep{auchettl18}. The colour bar on the right indicates the phase relative to the UV/optical peak, with darker colours indicating earlier times.}
 \label{fig:hr_lum}
\end{figure}

During the evolution of ASASSN-19dj, the HR and the X-ray luminosity seem to follow an inverse relationship, where the X-ray emission becomes harder as the luminosity of the source fades, and becomes softer as the source brightens. This relationship is shown in Figure \ref{fig:hr_lum}, with colour-coding and arrows to highlight the trend. This evolution is consistent with what is seen in highly variable, X-ray bright AGN \citep[c.f. Figure 4 of][]{auchettl18}. Additionally this behaviour is similar to the overall trends between X-ray luminosity and spectral hardness seen in the sample of \citet{wevers20}. The overall behaviour seen in ASASSN-19dj is quite unique compared to all other X-ray TDE candidates, even compared to ASASSN-15oi, which showed delayed brightening of the X-ray emission $\sim200$ days after peak UV brightness \citep{gezari17}, or ASASSN-18jd which showed large variations in HR with time before the emission completely faded \citep{neustadt19}. The correlated changes in HR and luminosity have not been seen before and the late-time brightening for ASASSN-19dj is different than that of ASASSN-15oi. \citet{auchettl18} showed that $<$4\% of X-ray bright AGN could produce flare emission that exhibits a coherent decay and a constant HR similar to that of an X-ray bright TDE. So while the brightening is similar to what we see from thermal TDEs such as ASASSN-14li, we cannot rule out that some of the emission arises from a pre-existing AGN disc \citep[e.g.][]{blanchard17}.

To further explore the nature of the X-ray emission arising from ASASSN-19dj, we analysed the \textit{Swift} and \textit{NICER} spectra using the X-ray spectral fitting program \textsc{XSPEC} version 12.10.1f \citep{arnaud96}, and chi-squared statistics. While we fit the majority of \textit{NICER} spectra individually (with the exception of a handful of observations at late times), it was necessary to stack the early-time \swift observations to get adequate S/N. We show the results of these spectral fits in the bottom three panels of Figure \ref{fig:all_xray}. 

At early times, the merged \swift spectrum is best fit by an absorbed blackbody plus power-law model. However, at late times, when ASASSN-19dj is significantly brighter, we find that an absorbed blackbody is sufficient to model the observed spectra. We let the column density ($N_{H}$), blackbody temperature ($kT$) and blackbody normalisation, as well as the photon-index $\Gamma$ and powerlaw normalisation for the early \swift spectra, of each model be free parameters. In Table \ref{tab:xray_spec} we summarise the best-fit parameters of our spectral fits. 

The third panel of Figure \ref{fig:all_xray} shows the column density as a function of time. The column densities derived using the \textit{NICER} spectra and the merged \swift spectra are all consistent with the Galactic column density along the line of sight, although the uncertainties are large. 

In Figure \ref{fig:all_xray} (fourth panel), we show the temperature evolution of ASASSN-19dj. We find that the derived X-ray blackbody temperatures are similar to other X-ray bright TDEs such as ASASSN-14li \citep{holoien16a, brown17a} and ASASSN-15oi \citep{gezari17, holoien18a}, and the very tail end of the blackbody temperature distribution of unobscured non-blazar AGN \citep{ricci17}, peaking at kT$\sim$ 110 eV. We find that the temperature of ASASSN-19dj is lower than derived for the TDE/AGN candidates ASASSN-18jd \cite{neustadt19} and ASASSN-18ul \citep{wevers19} which had blackbody temperatures more consistent with known AGN. Initially, we find that the blackbody component had a mean temperature of $\sim$60 eV. Unfortunately, due to the faintness of the source at early times, we are unable to constrain whether the temperature is constant with time or varies as seen during the late phases of its evolution. Interestingly, when the luminosity of the source increases after the seasonal gap, we find that the temperature initially does not change significantly from that seen at early times. However, as the source increases to peak brightness, we find that the temperature also increases, peaking at $62\pm5$ eV. As the source begins to fade, the temperature seems to follow the same short timescale variability behaviour seen in the X-ray light curve, suggesting that the short time scale luminosity variation we observe is dominated by changes in the blackbody temperature with time. The change in temperature is most dramatic between 200-280 days after the peak of the UV/optical light curve, where the blackbody temperature drops from $\sim$60 keV to $\sim$40 keV, before plateauing at this lower temperature value for the next $\sim$100 days, similar to what was seen in ASASSN-14li after peak X-ray brightness \citep[see Table 3 of][]{brown17a}.

In the bottom panel of Figure \ref{fig:all_xray} we show the evolution of the effective blackbody radius as a function of time. During its early evolution ASASSN-19dj has a blackbody radius that is consistent with ASASSN-18jd \citep{neustadt19} and ASASSN-15oi \citep{holoien18a} both before and after its observed X-ray brightening. The brightening of ASASSN-19dj after the seasonal gap is also associated with a dramatic order of magnitude increase in the blackbody radius. This suggests that the increase in X-ray luminosity is a result of an expansion of the X-ray emitting region rather than delayed accretion that could result from inefficient circularisation as suggested for ASASSN-15oi \citep[][]{gezari17} and previously suggested for ASASSN-19dj \citep{liu19}. \citet{vanvelzen20} also find that the X-ray brightening of ASASSN-19dj does not require delayed accretion. The size of the blackbody radius of ASASSN-19dj is consistent with ASASSN-14li \citep{brown17a}, ASASSN-15oi \citep{holoien18a}, and other TDE candidates whose blackbody radii were measured at peak using X-rays \citep[see Figure 11 of][]{2019MNRAS.487.4136W}. 

Similar to \citet{2019MNRAS.487.4136W} and individual studies of X-ray bright TDE candidates \citep[e.g.][]{brown17a, holoien18a}, we find that the X-ray emitting region is at least an order of magnitude smaller than the blackbody radius interfered from the UV/optical (see Figure \ref{fig:BB_fit}). Additionally, the blackbody radius is smaller than (at early times) and equal to (at peak brightness) the ISCO (innermost stable circular orbit) of the black hole, assuming a Swarzschild black hole. Due to the high cadence and high collecting area of \textit{NICER}, we are also able to observe how the blackbody radius changes as a function of time. We find that the blackbody radius does not vary significantly over short timescales and shows a slow overall increase with time. The slow evolution at late times is consistent with that seen in ASASSN-14li \citep{brown17a} and ASASSN-15oi \citep{holoien18a}.

\section{Discussion} \label{sec:disc}
ASASSN-19dj has one of the most complete datasets of any TDE to date. Its temporal coverage stretches from early in its evolution to more than a year after peak and spans the X-ray, NUV, and optical energy bands. As such, ASASSN-19dj provides us with the opportunity to make comparisons to the various theoretical models put forth to explain the UV/optical emission of TDEs. As described previously, the two main models are reprocessed disc emission \citep[e.g.][]{dai18, mockler19} and stream-stream collisions \citep[e.g.][]{jiang16, bonnerot17, lu20, ryu20a}. Since one of the major contentions of the reprocessing picture of TDE emission is that material surrounding the SMBH absorbs X-ray photons and re-emits them at longer wavelengths, our combination of a rich set X-ray and UV/optical observations is well suited to test these predictions.

A key feature of the evolution of ASASSN-19dj that may help distinguish between the models is the late-time X-ray brightening. \citet{liu19} suggest this is evidence for delayed accretion, similar to ASASSN-15oi \citep{gezari17}. ASASSN-19dj, however shows significant X-ray emission within days of the UV/optical peak. Given the deep upper limit on the X-ray emission from the first \swift epoch ($<$ 6$\times10^{40}$ erg s$^{-1}$) the early X-ray emission must be associated with the ASASSN-19dj TDE rather than any pre-existing AGN activity. This is evidence for prompt circularisation and the formation of an accretion disc. Additionally, the hardness-luminosity correlations shown in Figure \ref{fig:hr_lum} are consistent with variable accretion onto a compact object, further supporting the early-time creation of a disc.

Another possible explanation for the X-ray brightening is a changing obscuration, which allows us to see more of the dominant soft X-ray emission. There are two issues with this however. First, the stacked early-time \swift X-ray spectrum shows clear evidence of strong soft X-ray emission. Second, the column density inferred from the early-time \swift spectrum shows no evidence of being elevated relative to the column density at late times. The column densities are modest ($\sim5\times 10^{20}$ cm$^{-2}$) and consistent with Galactic.

Our dense spectral coverage of ASASSN-19dj also allow us to place this source in the broader context of TDE flares. The broad feature near 4600 \AA \ is commonly associated with Bowen fluoresence \citep{leloudas19, neustadt19, vanvelzen20} and requires FUV flux for creation. In the spectral sequence of ASASSN-19dj, this broad complex of lines is present in the first pre-peak spectrum and persists until the seasonal break. However, as seen in Figure \ref{fig:bluespec}, the makeup of this feature is complex and evolves rapidly from a broad hump to individual broad lines as the TDE fades. If an accretion disc is the dominant producer of FUV radiation, this would suggest that an accretion disc has been formed very early on in the evolution of ASASSN-19dj. However, many of the emission lines do not peak until after the bolometric peak, with H$\alpha$ peaking roughly two months after peak UV/optical light. This delay may be explained if the emission is driven by stream-stream collisions \citep[e.g.][]{piran15, krolik16, ryu20a}. We also note that the line profiles are consistent with theoretical predictions for outflows, including a delay in the appearance of strong Balmer emission \citep{roth18}.

\section{Summary}\label{summary}
We have presented multiwavelength photometric and spectroscopic data of the tidal disruption event ASASSN-19dj. For the third time, we observe the initial optical rise of a TDE and find that it is again consistent with flux $\propto t^2$. ASASSN-19dj is among the most UV luminous TDEs yet discovered, with a peak absolute $UVW2$ magnitude of $-21.01 \pm 0.04$ mag, as well as being among the hottest. The evolution of the UV/optical emission of ASASSN-19dj is roughly consistent with that of other TDEs, following the trend that more luminous TDEs decay more slowly after peak. Our set of twenty-four spectra follow the evolution of this TDE from 9 days before peak until 327 days after, showing significant changes in both the continuum and H$\alpha$ emission in this time period. The peak H$\alpha$ emission is delayed by roughly 25 days from the UV/optical luminosity peak.

Through a search of archival CRTS photometry, we find a previous flare in the host galaxy roughly 14.5 years prior to ASASSN-19dj. While the quality of the CRTS images was poor, we were able perform image centroiding and find that the location of the previous outburst is consistent with the nucleus. Given the fact that KUG 0810+227 is a post-starburst galaxy, we would expect the TDE rate to be significantly higher in this galaxy than an average galaxy, allowing for the possibility of an earlier TDE. However, given the available data, the earlier flare could also be a luminous nuclear supernova or some other form of accretion flare from the SMBH.

In addition to being luminous in the UV and optical, ASASSN-19dj increased in X-ray luminosity near peak UV/optical light. After the 2019 seasonal gap, the X-ray luminosity was observed by \textit{XMM-Newton} slew, \swift and \textit{NICER} to have increased by an additional order of magnitude. The increase in X-ray luminosity appears to be a consequence of an increase in the area of the X-ray emitting region, while the short term variability and late-time decrease seen in the X-ray light curve arise from changes in the X-ray temperature. 

ASASSN-19dj is one of the few tidal disruption events with extensive multiwavelength photometric and spectroscopic data spanning from before peak to more than a year after. Even so, details on the emission mechanisms and spectral evolution are difficult to constrain. This indicates the importance of surveys like ASAS-SN and their ability to quickly find and confirm TDEs early so that similarly comprehensive data sets can be constructed.

\section*{Acknowledgements}
We thank the referee for helpful comments and suggestions that have improved the quality of this manuscript. We thank the \swift PI, the Observation Duty Scientists, and the science planners for promptly approving and executing our \swift observations. We thank Decker French for providing some of the data used in Figure \ref{fig:ew_bpt}. We thank Stephen Smartt and Kenneth Smith for helpful discussions on archival Pan-STARRS data. We also thank Jorge Anais Vilchez, Abdo Campillay, Yilin Kong Riveros, Nahir Muñoz-Elgueta, Natalie Ulloa for conducting Swope observations.

We thank the Las Cumbres Observatory and its staff for its continuing support of the ASAS-SN project. LCOGT observations were performed as part of DDT award 2019B-003 to EG. ASAS-SN is supported by the Gordon and Betty Moore Foundation through grant GBMF5490 to the Ohio State University, and NSF grants AST-1515927 and AST-1908570. Development of ASAS-SN has been supported by NSF grant AST-0908816, the Mt. Cuba Astronomical Foundation, the Center for Cosmology  and AstroParticle Physics at the Ohio State University, the Chinese Academy of Sciences South America Center for Astronomy (CAS- SACA), the Villum Foundation, and George Skestos. 

BJS, CSK, and KZS are supported by NSF grant AST-1907570/AST-1908952. BJS is also supported by NSF grants AST-1920392 and AST-1911074. CSK and KZS are supported by NSF grant AST-181440. KAA is supported by the Danish National Research Foundation (DNRF132). MJG was supported in part by the NSF grant AST-1815034 and the NASA grant 16-ADAP16-0232. M.A.T acknowledges support from the DOE CSGF through grant DE-SC0019323. Support for GP and JLP is provided in part by FONDECYT through the grant 1191038 and by the Ministry of Economy, Development, and Tourism's Millennium Science Initiative through grant IC120009, awarded to The Millennium Institute of Astrophysics, MAS. TAT is supported in part by NASA grant 80NSSC20K0531. D. A. Coulter acknowledges support from the National Science Foundation Graduate Research Fellowship under Grant DGE1339067. We acknowledge Telescope Access Program (TAP) funded by the NAOC, CAS, and the Special Fund for Astronomy from the Ministry of Finance.

The UCSC transient team is supported in part by NSF grant AST-1518052, the Gordon \& Betty Moore Foundation, the Heising-Simons Foundation, and by a fellowship from the David and Lucile Packard Foundation to R.J.F.

Parts of this research were supported by the Australian Research Council Centre of Excellence for All Sky Astrophysics in 3 Dimensions (ASTRO 3D), through project number CE170100013.

This publication makes use of data products from the Wide-field Infrared Survey Explorer, which is a joint project of the University of California, Los Angeles, and the Jet Propulsion Laboratory/California Institute of Technology, funded by the National Aeronautics and Space Administration.

The Pan-STARRS1 Surveys (PS1) and the PS1 public science archive have been made possible through contributions by the Institute for Astronomy, the University of Hawaii, the Pan-STARRS Project Office, the Max-Planck Society and its participating institutes, the Max Planck Institute for Astronomy, Heidelberg and the Max Planck Institute for Extraterrestrial Physics, Garching, The Johns Hopkins University, Durham University, the University of Edinburgh, the Queen's University Belfast, the Harvard-Smithsonian Center for Astrophysics, the Las Cumbres Observatory Global Telescope Network Incorporated, the National Central University of Taiwan, the Space Telescope Science Institute, the National Aeronautics and Space Administration under Grant No. NNX08AR22G issued through the Planetary Science Division of the NASA Science Mission Directorate, the National Science Foundation Grant No. AST-1238877, the University of Maryland, Eotvos Lorand University (ELTE), the Los Alamos National Laboratory, and the Gordon and Betty Moore Foundation.

Support for ATLAS observations and data products was provided by NASA grant NN12AR55G and 80NSSC18K0284.

The CSS survey is funded by the National Aeronautics and Space Administration under Grant No. NNG05GF22G issued through the Science Mission Directorate Near-Earth Objects Observations Program.  The CRTS survey is supported by the U.S.~National Science Foundation under grants AST-0909182.

Some of the data presented herein were obtained at the W. M. Keck Observatory, which is operated as a scientific partnership among the California Institute of Technology, the University of California and the National Aeronautics and Space Administration. The Observatory was made possible by the generous financial support of the W. M. Keck Foundation.

The LBT is an international collaboration among institutions in the United States, Italy and Germany. LBT Corporation partners are: The University of Arizona on behalf of the Arizona Board of Regents; Istituto Nazionale di Astrofisica, Italy; LBT Beteiligungsgesellschaft, Germany, representing the Max-Planck Society, The Leibniz Institute for Astrophysics Potsdam, and Heidelberg University; The Ohio State University, and The Research Corporation, on behalf of The University of Notre Dame, University of Minnesota and University of Virginia.

This work is based on observations made by ASAS-SN, ATLAS, Pan-STARRS, UH88, and Keck. We wish to extend our special thanks to those of Hawaiian ancestry on whose sacred mountains of Maunakea  and Haleakal\=a, we are privileged to be guests. Without their generous hospitality, the observations presented herein would not have been possible.

\section*{Data availability}
The data underlying this article are available in the article and in its online supplementary material.

\bibliographystyle{mnras}
\bibliography{bibliography}

\begin{table*}
\centering
 \caption{Spectroscopic observations of ASASSN-19dj}
 \label{tab:spec}
 \begin{tabular}{cccccc}
  \hline
  MJD & Date & Telescope & Instrument & Rest Wavelength Range & Exposure Time\\ 
  & & & & (\AA) & (s)\\
  \hline
  58539.2 & 2019 February 25.2 & ESO New Technology Telescope 3.58-m & EFOSC2 & 3555$-$9027 & 1$\times$300 \\
  58546.1 & 2019 March 4.1 & du Pont 100-in & WFCCD & 3717$-$9390 & 1$\times$900 \\
  58548.1 & 2019 March 6.1 & du Pont 100-in & WFCCD & 3717$-$9390 & 3$\times$900 \\
  58550.1 & 2019 March 8.1 & du Pont 100-in & WFCCD & 3717$-$9390 & 1$\times$900 \\
  58554.3 & 2019 March 12.3 & University of Hawaii 88-in & SNIFS & 3229$-$9489 & 3$\times$1800\\
  58555.1 & 2019 March 13.1  & Southern Astrophysical Research Telescope 4.1-m &  GHTS &  3913$-$8701  &   1$\times$600 \\
  58570.1 & 2019 March 28.1  & Southern Astrophysical Research Telescope 4.1-m &  GHTS &  3913$-$8701  &  1$\times$450\\
  58572.0 & 2019 March 30.0 & Magellan Clay 6.5-m & LDSS-3 & 3619$-$9048 & 1$\times$400\\
  58573.3 & 2019 March 31.3 & Keck I 10-m & LRISp & 3130$-$9781 & 4$\times$1800 \\
  58583.1 & 2019 April 10.1 & du Pont 100-in & WFCCD & 3717$-$9390 & 3$\times$900 \\
  58588.2 & 2019 Aprril 15.3  & Lick Shane Telescope 120-in & Kast &  3067$-$10039            & 1$\times$1545 (blue) 3$\times$500 (red) \\
  58604.3 & 2019 May 1.3 & Keck I 10-m   & LRIS & 4500$-$9709  & 1$\times$190 (blue) 1$\times$180 (red) \\
  58608.2 & 2019 May 5.2 & Lick Shane Telescope 120-in & Kast &  3067$-$10039   & 1$\times$1545 (blue) 3$\times$500 (red) \\
  58612.3 & 2019 May 9.3 & University of Hawaii 88-in & SNIFS & 3229$-$9489 & 2$\times$1800  \\
  58616.2 & 2019 May 13.2 & Lick Shane Telescope 120-in & Kast & 3067$-$10039 & 1$\times$1230 (blue) 2$\times$600 (red) \\
  58786.5 & 2019 October 30.5  & Lick Shane Telescope 120-in & Kast &  3067$-$10039   & 2$\times$1230 (blue) 4$\times$600 (red) \\ 
  58789.6 & 2019 November 2.6 & University of Hawaii 88-in & SNIFS & 3229$-$9489 & 2$\times$2100  \\
  58794.5 & 2019 November 7.5 & Lick Shane Telescope 120-in & Kast & 3067$-$10039 & 1$\times$1865 (blue) 3$\times$600 (red) \\ 
  58806.3 & 2019 November 19.3 &  Magellan Baade 6.5-m & IMACS & 4157$-$9195 & 4$\times$900 \\
  58811.3 & 2019 November 24.3 & du Pont 100-in & WFCCD & 3717$-$9390 & 1$\times$1200\\
  58816.5 & 2019 November 29.5 & University of Hawaii 88-in & SNIFS & 3229$-$9489 &  1$\times$3600\\
  58839.3 & 2019 December 22.3 & Large Binocular Telescope 8.4 m & MODS & 3130$-$9781 & 4$\times$1200 \\
  58854.2 & 2020 January 6.2 & Lick Shane Telescope 120-in  & Kast &   3067$-$10039 & 2$\times$1060 (blue) 3$\times$700 (red) \\
  58875.4 & 2020 January 27.4 & Keck I 10-m  & LRIS &  3063$-$9709  & 1$\times$900 (blue) 1$\times$888 (red) \\
  \hline
 \end{tabular}\\
\begin{flushleft}Modified Julian Day, calendar date, telescope, instrument, wavelength range, and exposure time for each of the spectroscopic observations obtained of ASASSN-19dj for the initial classification and during our follow-up campaign.\end{flushleft}
\end{table*}

%Yep, ASAS-SN still rules!

\bsp
\label{lastpage}
\end{document}